\documentclass[titlepage,12pt]{utarticle}
\usepackage{amssymb, amsmath, amsopn, amsthm,amscd,amsfonts}
\usepackage{color}
\usepackage{latexsym}
\usepackage{ifthen}
\usepackage{epsfig}
\usepackage{graphicx}
\usepackage{graphics}
\usepackage{bm,longtable,enumerate}
\numberwithin{equation}{section}
\newcommand{\tr}{{\rm tr}}
\newcommand{\ic}{{\rm i}}

\newcommand{\be}{\begin{equation}}
\newcommand{\ee}{\end{equation}}
\newcommand{\bea}{\begin{eqnarray}}
\newcommand{\eea}{\end{eqnarray}}
\renewcommand{\d}{\mathrm{d}}
\newcommand{\e}{\varepsilon}

\def\eps{\epsilon}

\newcommand{\eqn}{\begin{eqnarray}}
\newcommand{\eqnx}{\end{eqnarray}}

\def\beq{\begin{equation}}
\def\eeq{\end{equation}}
\def\beqa{\begin{eqnarray}}
\def\eeqa{\end{eqnarray}}

\begin{document}
\preprint{{\tt arXiv:1010.0993}\\}
\title{{\bf Scalar and vector mesons of \vskip 0.1cm flavor chiral symmetry breaking \vskip 0.3cm in the Klebanov-Strassler background}}

\author{{\sc Matthias Ihl}, {\sc Marcus A. C. Torres}, {\sc Henrique Boschi-Filho}, 
 \address{
 Instituto de F{\'i}sica, \\
 Universidade Federal do Rio de Janeiro,\\
         Caixa Postal 68528,\\
         21941-972 Rio de Janeiro, RJ, Brasil.\\[0.2cm]
         Email: {\tt msihl@if.ufrj.br},\\
	 ~~~~~~~~~{\tt mtorres@if.ufrj.br},\\
	 ~~~~~~~~~{\tt boschi@if.ufrj.br}\\
	 }
\\ {\sc C. A. Ballon Bayona}
 \address{
  Centro Brasileiro de Pesquisas F{\'i}sicas,\\
         Rua Dr. Xavier Sigaud 150, Urca, 22290-180 Rio de Janeiro, RJ, Brasil,\\[0.2cm]
	 ~and\\[0.2cm]
	 Centre for Particle Theory, \\
         University of Durham, \\
	 Science Laboratories, South Road,\\
         Durham DH1 3LE, United Kingdom.\\[0.2cm]
         Email: {\tt ballon@cbpf.br}\\
}
}

\Abstract{Recently, Dymarsky, Kuperstein and Sonnenschein constructed an embedding of flavor D7- and anti-D7-branes in the Klebanov-Strassler
geometry that breaks the supersymmetry of the background, yet is stable. In this article, we study in detail the spectrum of vector mesons 
in this new model of flavor chiral symmetry breaking and commence an analytical study of the scalar mesons in this setup. }

\maketitle
\tableofcontents
\newpage

\section{Introduction}
The AdS/CFT correspondence \cite{Maldacena:1997re, Witten:1998qj} and its many non-conformal generalizations (often referred to as gauge/gravity or gauge/string dualities; for a recent review, see \cite{Gubser:2009md}, and references therein) have become promising and powerful tools to study non-perturbative physics of strongly coupled gauge theories in general and QCD in particular. There have been applications in the field of heavy ion physics, where experiments at RHIC have revealed that the QCD plasma is strongly coupled and close to a perfect quantum liquid (for a recent discussion see \cite{Sharma:2009zz}). The gauge/gravity correspondence can be applied and help elucidate the strongly coupled particle and hadronic physics of the quark gluon plasma \cite{Policastro:2001yc,Policastro:2002se,Policastro:2002tn}.
Another subject where gauge/gravity techniques have recently been sucessfully applied, are strongly coupled condensed matter systems (cf.~\cite{Hartnoll:2009sz,Herzog:2009xv} and references therein), such as quantum critical systems and high temperature superconductors \cite{Gubser:2008px,Gubser:2008zu,Gubser:2008wv}.\\
One of the best studied examples in the context of top-down\footnote{There are also many phenomenological models of QCD, inspired by holographic ideas. These bottom-up models are generically known as AdS/QCD, see, e.g., \cite{BoschiFilho:2002zs,BoschiFilho:2002ta,BoschiFilho:2002vd,Erlich:2005qh,Gursoy:2007cb}.} gauge/string models is the confining, non-conformal background with constant dilaton introduced by Klebanov and Strassler \cite{Klebanov:2000hb}, which is dual to a $3+1$-dimensional $\mathcal{N}=1$ supersymmetric field theory with gauge group $SU(N+M) \times SU(N)$. This field theory arises as the near-brane limit of $N$ $D3$-branes and $M$ wrapped $D5$-branes at the tip of the conifold. It exhibits logarithmic running of the gauge couplings which, along the RG flow, undergo a cascade of Seiberg dualities, where $N$ repeatedly decreases by $M$ units. In the far IR, the duality cascade must stop and the IR dynamics removes the conifold singularity via a gluino condensate that breaks the exact anomaly-free R-symmetry $\mathbb{Z}_{2M}$ down to $\mathbb{Z}_2$. We are left with a warped deformed conifold whose deformation parameter $\eps^2$ is related to the expectation value of the gluino condensate. The gauge theory dual, in the far IR, is in the same universality class as $\mathcal{N}=1$ $SU(M)$ SYM theory, albeit with extra massive adjoint scalars and fermions (for a review, see e.g., \cite{Gwyn:2007qf}).\\
The Klebanov-Strassler model and its generalizations are particularly interesting phenomenologically, since they can be utilized in many ways. An important role is played by the $D7$-branes in this context. Besides providing a holographic realization of QCD-like flavor physics of mesons, which will be the main focus of this paper, and baryons, it is also a crucial ingredient in string model building, i.e., compactifications with warped throats \cite{Kachru:2003aw, Burgess:2003ic}. Furthermore, it seems conceivable that cosmological inflation consistent with present observational data can be realized in such models \cite{Kachru:2003sx,Baumann:2006th,Baumann:2007ah}. The existence of meta-stable vacua has been investigated in \cite{Bena:2009xk}. Spectra of glueballs in the deformed conifold theory have been obtained in \cite{Caceres:2000qe} and in \cite{Mia:2009wj} a finite temperature version of the Klebanov-Strassler model was used to study the physics of the quark-gluon plasma.\\
From a theoretical point of view, much progress has been made over the past years, e.g., in connecting the Klebanov-Strassler background to the Gopakumar-Vafa open/closed string conjecture (see \cite{Gwyn:2007qf} for a review, and references therein). Moreover, its baryonic branch has been related to SU(3) structure backgrounds (\cite{Butti:2004pk}, also see \cite{Dymarsky:2005xt,Maldacena:2009mw, Gaillard:2010qg} for some recent developments) and the resolved warped deformed conifold. Its mesonic branch has been studied in \cite{Krishnan:2008gx}.\\
Recently, Dymarsky, Kuperstein and Sonnenschein \cite{Dymarsky:2009cm} were able to construct a non- supersymmetric embedding of $D7$- and $\overline{D7}$-branes that constitutes a geometric realization of flavor chiral symmetry breaking in the Klebanov-Strassler model. This is based on the 
original ideas by Sakai and Sugimoto \cite{Sakai:2004cn,Sakai:2005yt} and on similar results by Kuperstein and Sonnenschein \cite{Kuperstein:2008cq} obtained previously for the Klebanov-Witten model \cite{Klebanov:1998hh}. In an early attempt \cite{Sakai:2003wu} to incorporate a configuration of flavor $D7$-$\overline{D7}$-branes into the Klebanov-Strassler model, that in fact predates the Sakai-Sugimoto model, the full solution could not be obtained due to complications arising from the non-trivial NS-NS flux in the Klebanov-Strassler background. These problems have been overcome in \cite{Dymarsky:2009cm} by generalizing the results in \cite{Kuperstein:2008cq}.\\
In a recent paper \cite{Bayona:2010bg}, the authors studied phenomenological aspects of pions and vector mesons in the Kuperstein-Sonnenschein holographic model of chiral symmetry breaking based on the Klebanov-Witten background. The numerical results for the form factors, particularly for the pions, indicated that the meson physics in that model very well reproduces experimental data and QCD results.  
It is therefore quite natural to study phenomenological aspects of the Dymarsky-Kuperstein-Sonnenschein model, which is dual to a (large $N_c$) QCD-like theory in the IR, rather than a quiver gauge theory with gauge group $SU(N) \times SU(N)$, since it is based on the Klebanov-Strassler background.\\
The focus of the present article is the investigation of vector and axial-vector mesons (including pions), scalar mesons and their respective spectra within the framework of the DKS $D7/\overline{D7}$-brane embedding in the Klebanov-Strassler background. 
The paper is organized as follows: In order to be self-contained, we review the construction of Dymarsky-Kuperstein-Sonnenschein in section \ref{sec:DKS}. 
Vector and scalar meson spectra are investigated in detail in section \ref{sec:mesons}. While we find numerically robust results for the vector mesons, 
we did not find a numerically stable eigensystem for the scalar mesons, thus we only present analytical results on the corresponding Sturm-Liouville problem. 
We conclude with some remarks and an outlook in section \ref{sec:conclusions}.

\section{Review of the Dymarsky-Kuperstein-Sonnenschein model}\label{sec:DKS}
The Klebanov-Strassler background is a special case of the so-called imaginary self-dual (ISD) backgrounds in type IIB superstring theory with constant dilaton $e^{\Phi}=g_s$. We will first review the strategy for constructing non-supersymmetric $D7$-$\overline{D7}$-brane profiles for the general case of ISD backgrounds before specializing to the Klebanov-Strassler case.
\subsection{Non-BPS D7-brane embeddings in ISD backgrounds}
\noindent
{\bf ISD condition.} Let us briefly recall the definition of ISD backgrounds of type IIB superstring theory with constant dilaton $e^{\Phi}=g_s$ \cite{Grana:2000jj,Gubser:2000vg}. Such a ten dimensional background is 
a warped product 
\begin{equation}\label{eq:warp}
\d s_{(10)}^2 = h^{-1/2} \d x_{\mu} \d x^{\mu} + h^{1/2} \d s_{(6)}^2,
\end{equation}
of flat Minkowski space-time $M^{3,1}$ and an (internal\footnote{In mathematics, Calabi-Yau manifolds are understood to be compact K\"ahler manifolds with a Ricci-flat metric. In physics, one often allows for non-compact examples, such as the (singular) conifold, which is a cone over the base $T^{1,1}$. It is possible, however, to compactify the (warped deformed) conifold, i.e., embed it as a ``warped throat`` into a compact Calabi-Yau manifold.}) Calabi-Yau manifold $Y^6$ with metric $\d s_{(6)}^2$, where the warp factor $h$ may only 
depend on the coordinates on $Y^6$. The self-dual R-R 5-form is given by
\begin{equation}
\widetilde{F}_{(5)} = F_{(5)} + B_{(2)} \wedge F_{(3)} = (1+ \star_{10}) \d C_{(4)}, 
\end{equation}
where $F_{(3)}=\d C_{(2)}$, $F_{(5)} = \d C_{(4)}$ and $C_{(4)} = g_s^{-1} h^{-1} \d x^0 \wedge \d x^1 \wedge \d x^2 \wedge \d x^3$. 
The complex 3-form flux $G_{(3)} = F_{(3)} + \frac{\ic}{g_s} H_{(3)}$ is imaginary self-dual w.r.t. the metric on $Y^6$,  
\begin{equation}\label{eq:ISD}
 G_{(3)} = \ic \star_6 G_{(3)},
\end{equation}
and vanishes in the Minkowski directions. It was shown in \cite{Grana:2000jj} that if $G_{(3)}$ is of type $(2,1)$, then the background preserves $\mathcal{N}=1$ supersymmetry with four supercharges.\\
Let us consider the R-R form $C_{(6)}$. Its field strength is defined as
\begin{equation}
 F_{(7)} = \star_{10} F_{(3)} - C_{(4)} \wedge H_{(3)},
\end{equation}
with $\star_{10} F_{(3)}= h^{-1} \star_6 F_{(3)} \wedge \d x^0 \wedge \d x^1 \wedge \d x^2 \wedge \d x^3$.
Note that the equation of motion for $C_{(2)}$ implies that $\d F_{(7)}=0$. 
Using the explicit form of $C_{(4)}$ given above, and observing that the ISD condition on $G_{(3)}$ (\ref{eq:ISD}) implies
\begin{equation}
F_{(3)}= - g_s^{-1} \star_6 H_{(3)}, \quad \mathrm{and} \quad \star_6 F_{(3)}=  g_s^{-1} H_{(3)},
\end{equation}
it follows that $F_{(7)} =  0$ \cite{Kuperstein:2004hy}. Therefore it is consistent to set $C_{(6)}=0$ for the rest of the discussion.\\
\noindent
{\bf D7-brane embeddings in ISD backgrounds.} In this setup, $D7$-branes extend along the Minkowski directions and wrap a 4-cycle $\Sigma_4$ inside the internal space $Y^6$. Finding a classical $D7$-brane configuration entails solving the equations of motion for the scalars parametrizing the embedding and the field equations for the gauge fields on the worldvolume of the brane. Finding a solution is difficult and untractable in general; the situation is made even worse by the existence of a non-trivial NS-NS field $H_{(3)}= \d B_{(2)}$ in the background. A substantial simplification occurs if one is after an embedding preserving some supersymmetry. Scrutinizing the $\kappa$-symmetry of the theory, the authors of \cite{Becker:1995kb,Marino:1999af} showed that the problem reduces to finding a Euclidean $D3$-brane wrapping the 4-cycle $\Sigma_4$. More specifically, in order to preserve the supersymmetries of the background, the $D7$-brane embedding and the gauge field living on its worldvolume have to satisfy the following conditions:
\begin{itemize}
 \item The 4-cycle $\Sigma_4$ is holomorphic, i.e., $\Sigma_4$ is the zero locus of an equation that can be written purely in terms of holomorphic 
 variables\footnote{See appendix \ref{ap:A} for a discussion of different coordinate systems used to describe the warped deformed conifold.}.
 \item The gauge-invariant field strength $\mathcal{F}= P[B_{(2)}] + 2 \pi \alpha' F$ is of type $(1,1)$ and anti-selfdual (ASD) for $D7$-branes (selfdual (SD) for $\overline{D7}$-branes), 
                   $$\mathcal{F}= -  \star_4 \mathcal{F}, \qquad \left( \mathcal{F}= + \star_4 \mathcal{F} \right).$$
\end{itemize}
Unfortunately, supersymmetric embeddings of $D7$- and $\overline{D7}$-branes can be shown to be rather trivial in the UV, namely the solutions exhibit only a single branch. Therefore they are not useful for modelling flavor chiral symmetry breaking.\\ 
One of the main achievements of \cite{Dymarsky:2009cm} was the following observation:\\\\
{\sl Given any induced worldvolume metric $g_{(8)}$, the corresponding $D7$-brane ($\overline{D7}$-brane) action is bounded from below and minimized for $\mathcal{F}$ anti-selfdual (selfdual). The resulting configuration is stable.}\\\\
Stated differently, the equations of motion for the $D7$-brane are solved by any embedding that extremizes the volume together with any ASD gauge field. The two questions are independent of each other and can be addressed separately. This leads to the realization that not all ASD solutions are supersymmetric:\\\\
{\sl If the embedding has to satisfy boundary conditions (e.g., invariance under a certain symmetry group) which are incompatible
with holomorphicity, but minimizes the volume $\int_{\Sigma_4} \sqrt{|g_{(4)}|}$, for a certain induced $g_{(4)}$ within this class of geometries, then the resulting ASD solution is non-supersymmetric (sometimes also called non-BPS) and stable, since it minimizes the action.}\\\\
With this recipe in mind, we will now review the construction of non-BPS $D7$-$\overline{D7}$ configurations in the Klebanov-Strassler background.

\subsection{Non-BPS D7-brane embeddings in the Klebanov-Strassler background}
\begin{figure}[ht] 
\begin{center}
\epsfig{file=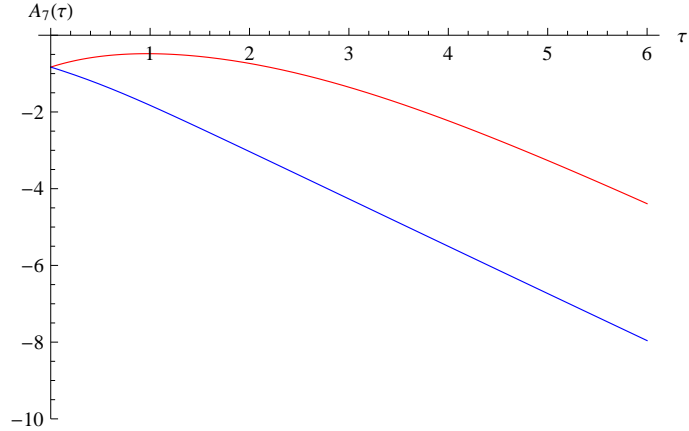, width=9cm}
\end{center}
\caption{The background gauge field for $\tau_0=0$: The ASD solution $A_7(\tau)$ on the $D7$-brane branch (blue), and the SD solution $\overline{A}_7(\tau)$ on the $\overline{D7}$-brane branch (red). The constant $c_1$ is chosen such that the two solutions match smoothly at the origin.}
\label{fig:A7}
\end{figure}

\noindent
{\bf The Klebanov-Strassler model.} Klebanov and collaborators successfully generalized the theory of $N$ $D3$-branes sitting at the tip of a singular conifold (the Klebanov-Witten model \cite{Klebanov:1998hh}) to include $M$ fractional $D3$-branes ($D5$ branes wrapped on the collapsed 2-cycle of the conifold). In a sequence of articles \cite{Klebanov:1999rd,Klebanov:2000nc}, culminating in \cite{Klebanov:2000hb}, they developed an understanding of the supergravity solution and its dual gauge theory. 
The ten dimensional supergravity solution is a warped deformed conifold (\ref{eq:warp}), where $h$ solely depends on the dimensionless radial coordinate $\tau$.
The six dimensional metric on the deformed conifold is given by \cite{Butti:2004pk} (see also \cite{Papadopoulos:2000gj}),
\begin{equation}\label{eq:6dmetric}
\d s_{(6)}^ 2 = \frac{\eps^{4/3}}{4} K(\tau) \cosh \tau \left [ e_1^2 + e_2^2 + \e_1^2 + \e_2^2 + \frac{2}{\cosh \tau} (e_1 \e_1 + e_2 \e_2) + \frac{2}{3 K(\tau)^3 \cosh \tau} (\d \tau^2 + \widetilde{\e}_3^2 ) \right ],
\end{equation}
where the vielbeins $e_i$, $i=1,2$ on $S^2$ and $\e_j$, $j=1,2,3$ on $S^3$ are defined in appendix \ref{ap:A}. Note that $\widetilde{\e}_3= \e_3 + \cos \theta_1 \d \phi_1$. Moreover,
\begin{equation}
K(\tau) = \frac{(\sinh \tau \cosh \tau - \tau)^{1/3}}{\sinh \tau}. 
\end{equation}
The warp factor can be obtained by solving a six-dimensional Laplace equation, yielding 
\begin{equation}
h(\tau) = (g_s M \alpha')^2 2^{2/3}\eps^{-8/3} \int_{\tau}^{\infty} \d x \frac{x\coth x - 1}{\sinh^2 x} \left(\sinh 2x - 2x\right)^{1/3}.
\end{equation}
The $D5$-branes wrapped over the $S^2$ of the $T^{1,1}$ are sources for magnetic R-R 3-form flux through the $S^3$ of $T^{1,1}$. Thus, in addition to $N$ units of 5-form flux, the solution also has $M$ units of 3-form flux, i.e.,
\begin{equation}
 \int_{S^3} F_{(3)} = M, \qquad \int_{T^{1,1}} F_{(5)} = N.
\end{equation}
In the following the NS-NS Kalb-Ramond field
\begin{equation}
B_{(2)} = b(\tau) \left[\e_1 \wedge \e_2 + e_1 \wedge e_2 + \frac{1}{\cosh \tau} (e_1 \wedge \e_2 + \e_1 \wedge e_2) \right],
\end{equation}
with 
\begin{equation}
 b(\tau) = -\frac{1}{2} g_s M \alpha' \frac{\cosh \tau (\tau \cosh \tau- \sinh \tau)}{2 \sinh^2 \tau},
\end{equation}
will play a crucial role.\\
The dual gauge theory is a $\mathcal{N}=1$ $SU(N+M) \times SU(N)$ gauge theory with two chiral superfields $A_{1,2}$ in the bifundamental $(N+M,\overline{N})$ color representation and two chiral superfields $B_{1,2}$ in the bifundamental $(\overline{N+M},N)$ color representation whose superpotential reads $\mathcal{W}= \lambda \e^{ik} \e^{jl} A_i B_j A_k B_l$. The theory is invariant under a $SU(2) \times SU(2) \times U(1)$ global symmetry and a discrete $\mathbb{Z}_{2M}$ R-symmetry. 
The relative gauge coupling 
\begin{equation}
g_1^{-2}-g_2^{-2} \sim g_s^{-1} \left(\int_{S_2} B_{(2)} -\frac{1}{2} \right), \quad \int_{S_2} B_{(2)} \sim M g_s \ln (r/r_{\mathrm{UV}}),
\end{equation}
runs logarithmically due to the dependence of $B_{(2)}$ on the radial coordinate. 
Here, $r_{\mathrm{UV}}$ is an arbitrary UV scale, where it should be noted that $r^2 \sim \rho^{4/3}$ for large radial distances, with $\rho^2= \eps^2 \cosh \tau$ (see appendix \ref{ap:A}).
Thus, the magnetic 3-form flux is responsible for the breaking of conformal symmetry in this model. The self-dual 5-form flux becomes
\begin{equation}
\tilde{F}_{(5)} = \left(N+ a g_s M^2 \ln (r/r_{\mathrm{UV}})\right) \mathrm{vol}\, T^{1,1}, 
\end{equation}
where $\mathrm{vol}\, T^{1,1}$ is the volume form of the conifold base $T^{1,1}$ and $a$ is a constant of order one.\\
A discussion of the Seiberg duality cascade, both from the points of view of the gauge theory and the supergravity can
be found in the original article \cite{Klebanov:2000hb} (cf.~also \cite{Dymarsky:2005xt}). 
There is an RG cascade, in which the 5-form flux $\tilde{F}_{(5)}$, present at some UV scale $\tau_{UV}$, repeatedly drops by $M$ units as $\int_{S^2} B_{(2)}$ goes through a period, ultimately vanishing in the IR. The important observation is that this cascade can be related on the dual gauge theory side to a cascade of $\mathcal{N}=1$ transformations on the gauge group factors, so-called Seiberg dualities. Since $g_1$ and $g_2$ flow in opposite directions, these duality transformations have to be performed at scales $\Lambda_i$ where one of the coupling constants diverges. Obeying certain matching conditions, it turns out that the new theory has the same form as the old theory under $N \rightarrow N-M$, thus resulting in a self-similar RG flow. When the cascade stops in the far IR, there are no $D3$-branes left other than the $M$ fractional $D3$-branes, and the gauge theory is $\mathcal{N}=1$ $SU(M)$ SYM. The conifold singularity is removed by blowing up the $S^3$ of $T^{1,1}$, as can be seen by studying the resulting geometry via introducing a probe $D3$-brane: The probe brane lives on a deformed conifold.\\
\noindent
{\bf Non-supersymmetric D7-brane embeddings.} We can now apply the strategy outlined for the general case above to the case at hand. Recall that the $D7$-branes extend in the flat Minkowski directions. The 4-cycle $\Sigma_4$ describing the embedding on the deformed conifold necessarily extends along the radial direction $\tau$. Moreover, in order to be $SU(2)_R$ invariant, $\Sigma_4$ must cover the $S^3$ of the $T^{1,1}$ base completely. Therefore, the $D7$-branes are located at points in $S^2$ that depend on $\tau$, similar to the situation in \cite{Kuperstein:2008cq}. Hence, the embedding describes a trajectory $(\phi_1(\tau),\theta_1(\tau))$ on $S^2$. Without loss of generality, one can set $\theta_1 = \pi/2$. The resulting induced metric $g_{(8)}=P[g_{(10)}]$ is the pullback of the ten-dimensional metric to the world volume of the $D7$-brane and reads
\begin{equation}\label{eq:D7metric}
 \d s_{(8)}^2 = h^{-1/2} (\tau) \d x_{\mu} \d x^{\mu} + h^{1/2} (\tau) \d s_{\Sigma_4}^2,
\end{equation}
where 
\begin{equation}
\eps^{-4/3} \d s_{\Sigma_4}^2 = \frac{K(\tau)}{4}  \cosh \tau \left[ \left(\tanh^2 \tau \left(\frac{\partial \phi_1}{\partial \tau}\right)^2 + \frac{2}{3 K(\tau)^3 \cosh \tau} \right) \d \tau^2 + \e_1^2 + \widetilde{\e}_2^2  + \frac{2}{3 K(\tau)^3 \cosh \tau} \e_3^2 \right],
\end{equation}
with $\tilde{\e}_2:= \e_2 + \frac{\partial \phi_1/\partial \tau}{\cosh \tau} \d \tau$.
The classical configuration that solves the equation of motion can be determined from the DBI action of the $D7$-branes without gauge fields, integrated over the Minkowski and $S^3$ directions,
\begin{equation}\label{eq:DBI0action}
S^{(0)}_{\text{DBI}} =- \mu_7 g_s^{-1} \frac{2\pi^2}{24} \eps^{8/3} \text{Vol}_{M^{3,1}} \int_0^{\infty} \d \tau \frac{\cosh \tau}{K(\tau)} 
\left( 1+ \frac{3 K(\tau)^3 \sinh^2 \tau}{2 \cosh \tau} \left(\frac{\partial \phi_1}{\partial \tau}\right)^2\right)^{1/2} 
\end{equation}
yielding
\begin{equation}\label{eq:Ushape}
 \left(\frac{\partial \phi_1}{\partial \tau}\right)^2 = \frac{2 \cosh \tau}{3 K(\tau)^3 \sinh^2 \tau} \left( \frac{K(\tau) \sinh^2 \tau \cosh \tau}{K(\tau_0) \sinh^2 \tau_0 \cosh \tau_0}-1\right)^{-1}.
\end{equation}
This leads to the U-shaped configuration that is needed for a geometric realization of flavor chiral symmetry breaking: The $D7$-brane profile stretches down to a minimal radius $\tau=\tau_0$, where $\frac{\partial \phi_1}{\partial \tau}$ jumps from $- \infty$ to $+ \infty$, i.e., the $D7$-brane turns around at $\tau=\tau_0$, becoming a $\overline{D7}$-brane.
The minimal radial distance from the tip of the deformed conifold, $\tau_0\geq 0$, parametrizes a one-parameter family of solutions to the embedding equations.\\
We can calculate the total angle $\Delta \phi_1$ covered by the trajectory on $S^2$ by integrating $\frac{\partial \phi_1}{\partial \tau}$,
\begin{equation}
\frac{1}{2} \Delta \phi_1 = \left(\frac{2}{3} \right)^{1/2} \int_{\tau_0}^{\infty} \d \tau \left(\frac{\cosh \tau}{K(\tau)^3 \sinh^2 \tau} \right)^{1/2} \left(\frac{K(\tau) \sinh^2 \tau \cosh \tau}{K(\tau_0) \sinh^2 \tau_0 \cosh \tau_0}-1\right)^{-1/2}. 
\end{equation}
There are two interesting limits: 
For small $\tau$, $K(\tau)$ approaches $(2/3)^{1/3}$, thus in the limit $\tau_0 \rightarrow 0$, the total angle evaluates to
\begin{equation}
 \frac{1}{2} \Delta \phi_1 (\tau_0 \rightarrow 0) = \int_{\tau_0}^{\infty} \frac{\d \tau}{\tau \left( \frac{\tau^2}{\tau_0^2}-1\right)^{1/2}} = \frac{\pi}{2}.
\end{equation}
This corresponds to an antipodal configuration.\\
On the other hand, if $\tau$ is large, $K(\tau) \approx 2^{1/3} e^{-\tau/3}$, and thus in the large $\tau_0$ limit, one obtains
\begin{equation}
\frac{1}{2} \Delta \phi_1 (\tau_0 \rightarrow \infty) = \frac{\sqrt{6}}{8} \pi,
\end{equation}
which exactly reproduces the result found in \cite{Kuperstein:2008cq} in the conformal Klebanov-Witten setup.
\noindent
{\bf ASD gauge field on the D7-branes.} To complete the construction of a non-supersymmetric $D7/\overline{D7}$-brane embedding, we need to find an (A)SD gauge field configuration $\mathcal{F}$ on its world volume. We will briefly reproduce here the results found in \cite{Dymarsky:2009cm}. One starts with writing down a gauge-invariant combination of the field strength on $\Sigma_4$ in the $A_{\tau}=0$ gauge, subject to the Bianchi identity,
\begin{equation}
\d \mathcal{F} = P[H_{(3)}], 
\end{equation}
with gauge field components on the $S^3$, $A_{\alpha=5,6,7}$. The pullback of the $B_{(2)}$ field to the 4-cycle $\Sigma_4$ is given by 
$P[B_{(2)}]=b(\tau) \e_1 \wedge \tilde{\e}_2$. Then the ansatz for the field strength reads
\begin{eqnarray}\label{eq:curlyF}
\mathcal{F} &=& P[B_{(2)}] +  F, \nonumber\\
 &=& \left(A_7'(\tau) + \frac{A_5(\tau)}{\cosh \tau} \phi_1'(\tau) \right) \d \tau \wedge \e_3 + \left(A_5'(\tau) - \frac{A_7(\tau)}{\cosh \tau} \phi_1'(\tau) \right) \d \tau \wedge \e_1 \nonumber\\
&& \qquad + A_6' \d \tau \wedge \e_2 - (A_7 (\tau) - b(\tau)) \e_1 \wedge \tilde{\e}_2 - A_5 (\tau) \tilde{\e}_2 \wedge \e_3 + A_6 (\tau) \e_1 \wedge \e_3,
\end{eqnarray}
where we have absorbed a dimensionful factor of $2\pi \alpha'$ into the gauge field components.
Here, the prime $'$ denotes derivation w.r.t. $\tau$. One should keep in mind that $\tau$ is only a good coordinate on each branch separately. Using the metric (\ref{eq:6dmetric}), we can write down the ASD condition for the $D7$-brane branch,
\begin{subequations}\label{eq:ASD}
\begin{eqnarray}
\left(A_5' (\tau) - \frac{A_7(\tau)}{\cosh \tau}\phi_1'(\tau) \right) &=& L_{\tau_0}(\tau) A_5 (\tau),\\
A_6' (\tau) &=& L_{\tau_0}(\tau) A_6 (\tau),\\
\left(A_7' (\tau) + \frac{A_5(\tau)}{\cosh \tau}\phi_1'(\tau) \right) &=& \frac{2 L_{\tau_0}(\tau)}{3 K(\tau)^3 \cosh \tau} \left( A_7 (\tau)-b(\tau)\right),
\end{eqnarray}
\end{subequations}
with
\begin{equation}
L_{\tau_0}(\tau) = \left(1+ \frac{3 K(\tau)^3 \sinh^2 \tau}{2 \cosh \tau}\phi_1'(\tau)^2\right)^{1/2}= \left(1- \frac{K(\tau_0) \sinh^2\tau_0 \cosh \tau_0}{K(\tau) \sinh^2\tau \cosh \tau}\right)^{-1/2}. 
\end{equation}
On the other branch, for the SD gauge fields on the $\overline{D7}$-branes, one has to replace $L_{\tau_0}(\tau) \rightarrow -L_{\tau_0}(\tau)$. It is also possible to switch to the globally well-defined coordinate $\phi(\tau)$ instead and reexpress the (A)SD equations in terms of $\phi$. 
It was shown in \cite{Dymarsky:2009cm} that $A_6 =  0$ to ensure regularity of the solution (otherwise $A_6 \propto e^{\tau}$ for large $\tau$). Unfortunately, there is no analytic solution for the remaining two gauge field components. \\
For $\tau_0=0$, one finds that the only non-divergent ASD solution is given by $A_5(\tau)=0$ and
\begin{equation}\label{eq:A7tau0}
A_7(\tau) = e^{S(\tau)} \left( \int_{\tau}^{\infty} \d \hat{\tau} \frac{2 b(|\hat{\tau}|)}{3 K(\hat{\tau})^3 \cosh \hat{\tau}}  e^{-S(\hat{\tau})} + c_0 \right),  
\end{equation}
where $c_0$ is an integration constant and 
\begin{equation}
S(\tau)= \int_0^{\tau} \d \hat{\tau} \frac{2}{3 K(\hat{\tau})^3 \cosh \hat{\tau}}.
\end{equation}
It is necessary to set $c_0=0$ to avoid an exponentially growing contribution to $A_7(\tau)$ for large $\tau$.
The non-divergent SD solution $\overline{A}_7(\tau)$ on the $\overline{D7}$-brane is
\begin{equation}
 \overline{A}_7(\tau) = e^{-S(\tau)} \left( \int_{0}^{\tau} \d \hat{\tau} \frac{2 b(|\hat{\tau}|)}{3 K(\hat{\tau})^3 \cosh \hat{\tau}}  e^{+S(\hat{\tau})} +c_1 \right),  
\end{equation}
where $c_1 \approx -0.83$ has to be adjusted such that the SD and ASD solutions are continuous at the origin $\tau=0$. 
We show the numerical solution for the background gauge field in figure \ref{fig:A7}.
For $\tau > \tau_0$, it can be shown that there is a unique solution which is regular both in the UV and at the tip of the profile $\tau=\tau_0$ (see appendix B of \cite{Dymarsky:2009cm} for details). Note that the solutions are uniquely fixed by the boundary conditions at $\tau=\tau_0$, namely  to ensure that the ASD gauge fields on the $D7$-brane branch are smoothly connected to the SD gauge fields on the $\overline{D7}$-brane branch.

\section{Mesons}\label{sec:mesons}

\subsection{Vector mesons}
According to the gauge/gravity correpondence, vector and axial-vector mesons correspond to fluctuations of the $U(N_f)$ gauge fields living on the $D7/\overline{D7}$-branes. The $D7$-brane action consists of two parts, 
\begin{eqnarray}\label{eq:D7action}
S_{D7} &=& S_{\text{DBI}} + S_{\text{WZ}}, \nonumber\\ &=& - \mu_7 \int d^4x d\tau d^3\Omega e^{-\Phi} \sqrt{-|g_{(8)} + \mathcal{F}|} 
- \mu_7 \int \sum_p C_{(p)} \wedge e^{\mathcal{F}} ,
\end{eqnarray}
where the first part is the Dirac-Born-Infeld action of the $D7/\overline{D7}$ branes and the second part is the Wess-Zumino action. The induced metric $g_{(8)}$ is given by (\ref{eq:D7metric}) and $\mathcal{F}$ was defined above.
For the case at hand, only the $C_{(4)}$ contributes to the Wess-Zumino action,
\begin{equation}
S_{\text{WZ}}= - \mu_7 \int \frac{1}{2} \mathcal{F} \wedge \mathcal{F} \wedge C_{(4)}. 
\end{equation}
Under the weak gauge field assumption, we can expand the DBI action as
\begin{equation}
S_{\text{DBI}}[g,\mathcal{F} + \delta \mathcal{F} ] = S^{(0)}_{\text{DBI}}[g] + S_{\text{YM}}[ (\delta \mathcal{F})^2 ] + \mathcal{O}[(\delta \mathcal{F})^4 ],
\end{equation}
where $S^{(0)}_{\text{DBI}}\sim \int_{\Sigma_4} \sqrt{|g_{(4)}|} $ was used above to determine the classical $D7$-brane embedding (\ref{eq:Ushape}). 
We want to expand this action around the classical action (\ref{eq:D7action}) and study a perturbation of the form $\delta \mathcal{F}= 2\pi \alpha' \hat F$, where the only non-vanishing components of the field strength $F$ are along the Minkowski directions and the radial direction $\tau$.
Following \cite{Dymarsky:2009cm}, we define
\begin{equation}
E := g_{(8)} + \mathcal{F} \; \Longrightarrow \sqrt{|E|} =  \sqrt{|g_{(4)}|} + h^{-1} \sqrt{|\mathcal{F}|},
\end{equation}
where $g_{(4)}$ is the induced metric on $\Sigma_4$. Its inverse is (schematically) given by
\begin{equation}\label{eq:Einv}
E^{-1} = \left(h^{1/2} \eta^{\mu \nu}, \frac{h^{-1/2} \sqrt{|g_{(4)}|}}{\sqrt{|g_{(4)}|} + h^{-1} \sqrt{|\mathcal{F}|}}
\left(g_{(4)}^{-1} - h^{-1/2} g_{(4)}^{-1} \mathcal{F} g_{(4)}^{-1} \right)\right) 
\end{equation}
The action $\delta S_{\text{YM}}\sim \mathcal{O}(\hat F^2)$ can then be integrated over the $S^3$ directions to give rise to the five-dimensional effective action for the gauge fields (for details, see appendix C of \cite{Dymarsky:2009cm}).
\begin{eqnarray}
S^{A_{\mu}}_{5d, \text{eff}}&=& -\kappa \int \d^4 x \d \tau \sqrt{|E|} \; \tr \left((E^{-1}\hat F)^2\right) \nonumber \\
&=& - \kappa \int \d^4 x \d \tau \left(\frac{1}{2} C_{\tau_0}(\tau)  \eta^{\mu \lambda} \eta^{\nu \rho} \hat F_{\lambda \rho} \hat F_{\mu \nu} + 
 D_{\tau_0}(\tau) \eta^{\mu \nu} \hat F_{\mu \tau} \hat F_{\nu \tau} \right),
\end{eqnarray}
where $\kappa = 2(\pi \alpha')^2 \mu_7 g_s^{-1} \text{Vol}\,S^3$. Here we have introduced 
\begin{eqnarray}
C_{\tau_0}(\tau) &=& h(\tau) \sqrt{|g_{(4)}|} + \sqrt{|\mathcal{F}|} \\ &=&\frac{\eps^{8/3}}{24}\frac{h(\tau) L_{\tau_0}(\tau)}{K(\tau)}\cosh \tau + A_5(\tau) \left(A_5'(\tau) -b(\tau) \frac{\phi_1'(\tau)}{\cosh \tau}\right) + \left(A_7(\tau)-b(\tau)\right) A_7'(\tau),\nonumber \\
D_{\tau_0}(\tau) &=& \sqrt{|g_{(4)}|} g_{(4)}^{\tau \tau} = \frac{\eps^{4/3}}{4}  \frac{K(\tau) \cosh \tau}{L_{\tau_0}(\tau)}.
\end{eqnarray}
Since $C_{\tau_0}(\tau)$ depends on $\mathcal{F}$, it is different for $D7$- and $\overline{D7}$-branes. Again, the two branches have to be smoothly connected at $\tau=\tau_0$.
For the study of meson spectra below, it turns out to be advantageous to use rescaled, dimensionless variants of the functions introduced above to explicitly introduce a mass scale\footnote{As explained in \cite{Klebanov:2000hb}, the four-dimensional mass scale is set by $M_{\text{gb}}\sim M_{\ast}\sim \frac{\eps^{2/3}}{g_s M \alpha'}$. This is the mass scale associated with the glueballs of the gauge theory. There is another independent mass scale associated with the mesons $M_{\text{meson}}\sim M_{\text{gb}} f(\tau_0)$, where $f(\tau_0)$ will be determined below. For $\tau_0=0$, the two mass scales are identical, $M_{\text{meson}}=M_{\text{gb}}$.} $M_{\ast}$:
\begin{eqnarray}\label{eq:5daction}
S^{A_{\mu}}_{5d, \text{eff}}&=&  - \hat{\kappa} \int \d^4 x \d \tau \left(\frac{1}{2} \widehat{C}_{\tau_0}(\tau) \eta^{\mu \lambda} \eta^{\nu \rho} \hat F_{\lambda \rho} \hat F_{\mu \nu} + M_{\ast}^2 \widehat{D}_{\tau_0}(\tau) \eta^{\mu \nu} \hat F_{\mu \tau} \hat F_{\nu \tau} \right),
\end{eqnarray}
where $\hat{\kappa}= \frac{(g_s M \alpha')^2}{24}\kappa$, $M_{\ast}^2= 6 \frac{\eps^{4/3}}{(g_s M \alpha')^2}$ and
\begin{eqnarray}
\widehat{C}_{\tau_0}(\tau) &=& \hat{h}(\tau) \frac{L_{\tau_0}(\tau)}{K(\tau)} \cosh \tau \nonumber \\
&& +  \hat{A}_5(\tau) \left(\hat{A}_5'(\tau) -\hat{b}(\tau) \frac{\phi_1'(\tau)}{\cosh \tau}\right) + \left(\hat{A}_7(\tau)-\hat{b}(\tau)\right) \hat{A}_7'(\tau),\\
\widehat{D}_{\tau_0}(\tau) &=& \frac{K(\tau) \cosh \tau}{L_{\tau_0}(\tau)}.
\end{eqnarray}
Here we have defined the dimensionless functions\footnote{Due to the different normalization of $b(\tau)$ compared to \cite{Dymarsky:2009cm}, we will have a slightly different asymptotics for $A_7(\tau)$, namely $A_7(\tau \rightarrow \infty) = \frac{1}{2} \sqrt{\frac{3}{2}}(-2 \tau -1)$, which differs by a factor of $6^{-1/2}$ from the DKS result.}
\begin{subequations}\label{eq:dimless}
\begin{eqnarray}
\hat{h}(\tau)&=&(g_s M \alpha')^{-2} \eps^{8/3} h(\tau) = 2^{2/3}\int_{\tau}^{\infty} \d x \frac{x\coth x - 1}{\sinh^2 x} \left(\sinh 2x - 2x\right)^{1/3},\\
\hat{b}(\tau)&=& \frac{2 \sqrt{6}}{g_s M \alpha'} b(\tau)= - \sqrt{6}\, \frac{\cosh \tau (\tau \cosh \tau- \sinh \tau)}{2 \sinh^2 \tau},\\
\hat{A}_{\alpha}(\tau) &=& \frac{2 \sqrt{6}}{g_s M \alpha'} A_{\alpha} (\tau).
\end{eqnarray}
\end{subequations}
From now on we will drop the hats, but it is to be understood that we will work solely with dimensionless quantities.
In the following, we want to study fluctuations of the gauge fields around the classical solution that correspond to vector mesons, axial-vector mesons and pions.
For convenience, we will work in the $A_{\tau}=0$ gauge. Then, $A_{\mu}$ can be expanded as
\begin{equation}\label{eq:Amuexp}
 A_{\mu}(x,\tau)= \widetilde{\mathcal{V}}_{\mu}(x)+ \widetilde{\mathcal{A}}_{\mu}(x) \psi_0(\tau) + \sum_{n=1}^{\infty} v_{\mu}^{(n)}(x) \psi_{2n-1}(\tau)+ \sum_{n=1}^{\infty} a_{\mu}^{(n)}(x) \psi_{2n}(\tau),
\end{equation}
where the zero-mode $\psi_0 (\tau)$ will be addressed later and the fields $v_{\mu}^{(n)}$ and $a_{\mu}^{(n)}$ 
are associated with vector and axial-vector mesons, respectively. 
In the $A_{\tau} =0$ gauge the pion field $\Pi(x)$ appears in the expansion of 
\begin{eqnarray}
 \widetilde{\mathcal{V}}_{\mu}(x) &=& \frac{1}{2}\left( A_{L\mu}^{\xi_+}(x) + A_{R\mu}^{\xi_-}(x)\right) \\\nonumber  
 &=&\frac{1}{2} \xi_+ \left( A_{L \mu} (x) + \partial_{\mu} \right)\xi_+^{-1} + \frac{1}{2} \xi_- \left( A_{R \mu} (x) + \partial_{\mu} \right)\xi_-^{-1},\\
 \widetilde{\mathcal{A}}_{\mu}(x) &=& \frac{1}{2}\left( A_{L\mu}^{\xi_+}(x) - A_{R\mu}^{\xi_-}(x)\right) \\\nonumber
 &=& \frac{1}{2} \xi_+ \left( A_{L \mu} (x) + \partial_{\mu} \right)\xi_+^{-1} - \frac{1}{2} \xi_- \left( A_{R \mu} (x) + \partial_{\mu} \right)\xi_-^{-1},
\end{eqnarray}
where $A_{L\mu}(x)$ and $A_{R\mu}(x)$ are external gauge fields and we have introduced $ \xi_+^{-1} = \xi_- := e^{i \frac{\Pi(x)}{f_\pi}}$. 
The normalization conditions and equations of motion for the wave functions $\psi_n (\tau)$ are determined to be,   
\begin{eqnarray}
\kappa \int d\tau \; C_{\tau_0}(\tau) \psi_m(\tau) \psi_n(\tau) &=& \delta_{m,n}, \label{eq:norm} \\
- (C_{\tau_0}(\tau))^{-1} \partial_{\tau} (D_{\tau_0}(\tau) \partial_{\tau} \psi_n (\tau)) &=& \lambda_n \psi_n(\tau),\label{eq:eom}
\end{eqnarray}
with $\lambda_n := M_n^2/M_{\ast}^2$. \\
\noindent
{\bf Four-dimensional effective action for the mesons.}
Substituting the gauge field (\ref{eq:Amuexp}) into the action $S_{5d, \text{eff}}$, eq.~(\ref{eq:5daction}), we find the four-dimensional effective Lagrangian for the vector and axial-vector mesons (disregarding divergent terms from non-renormalizable contributions), $\mathcal{L}_{4d, \text{eff}} = \mathcal{L}_{4d, \text{div}} + \sum_{j=2}^{\infty} \mathcal{L}_{4d, \text{eff}}^{(j)}$. The kinetic part of the Lagrangian reads
\begin{eqnarray}\label{eq:Lag2}
 \mathcal{L}_{4d, \text{eff}}^{(2)} &=& \frac{1}{2} \text{tr} (\partial_{\mu} \tilde{v}_{\nu}^{(n)} - \partial_{\nu} \tilde{v}_{\mu}^{(n)})^2 
 + \frac{1}{2} \text{tr} (\partial_{\mu} \tilde{a}_{\nu}^{(n)} - \partial_{\nu} \tilde{a}_{\mu}^{(n)})^2 + \text{tr} (i \partial_{\mu} \Pi + f_{\pi} \mathcal{A}_{\mu})^2 \nonumber \\
&& + M^2_{v^n}  \text{tr} \left( \tilde{v}_{\mu}^{(n)} - \frac{g_{v^n}}{M^2_{v^n}} \mathcal{V}_{\mu}\right)^2 + M^2_{a^n}  \text{tr} \left( \tilde{a}_{\mu}^{(n)} - \frac{g_{a^n}}{M^2_{a^n}} \mathcal{A}_{\mu}\right)^2,  
\end{eqnarray}
where, in order to diagonalize the kinetic terms, we have rewritten the meson fields in the following way:
\begin{equation}
\tilde{v}_{\mu}^{(n)} = v_{\mu}^{(n)}+\frac{g_{v^n}}{M^2_{v^n}}\mathcal{V}_{\mu}, \quad \tilde{a}_{\mu}^{(n)} = a_{\mu}^{(n)}+\frac{g_{a^n}}{M^2_{a^n}}\mathcal{A}_{\mu}. 
\end{equation}
Moreover, we define
\begin{eqnarray}
M_{v^n}^2 &=& \lambda_{2n-1} M_{\ast}^2, \quad M_{a^n}^2 = \lambda_{2n} M_{\ast}^2, \\
\mathcal{V}_{\mu} &=& \frac{1}{2} (A_{L\mu} + A_{R\mu}), \quad  \mathcal{A}_{\mu} = \frac{1}{2} (A_{L\mu} - A_{R\mu}).
\end{eqnarray}
The coupling constants between a massive vector meson $\tilde{v}_{\nu}^{(n)}$ (axial-vector meson $\tilde{a}_{\nu}^{(n)}$) and an external $U(1)$ field $\mathcal{V}_{\mu}$ representing a photon (an external axial $U(1)$ field $\mathcal{A}_{\mu}$) are
\begin{eqnarray}
g_{v^n} &=& \kappa M_{v^n}^2 \int_{D7/\overline{D7}} d\tau \,C_{\tau_0}(\tau) \psi_{2n-1} (\tau), \label{eq:gv}\nonumber \\
&=& -2 \kappa M_{\ast}^2 \left( D_{\tau_0}(\tau) \partial_{\tau} \psi_{2n-1}(\tau)\right)\Big{|}_{\tau\rightarrow \infty}, \\
g_{a^n} &=& \kappa M_{a^n}^2 \int_{D7/\overline{D7}} d\tau \, C_{\tau_0}(\tau) \psi_{2n} (\tau)\psi_0 (\tau).
\end{eqnarray}
In the next section, we will proceed to numerically calculate the wave functions $\psi_n(\tau)$ in order to be able to compute the mass spectrum 
$M_{v^n}^2$ and $M_{a^n}^2$.\\
\noindent
{\bf Study of the numerical wavefunctions.}
For $\tau_0=0$, the (extremal) $D7/\overline{D7}$-brane embedding touches the tip of the deformed conifold, where the $S^2$ shrinks to zero size. On the two branches the value of $\phi_1$ is constant, i.e., $\phi_1' =0$. At $\tau=0$, $\phi_1'$ diverges and the value of $\phi_1$ jumps from $- \frac{\pi}{2}$ to $+ \frac{\pi}{2}$. In this case it is possible to describe the $A_7$ component of the background gauge field on the $\overline{D7}$-brane branch by a simple continuation of (\ref{eq:A7tau0}) to negative values of $\tau$ (see \cite{Dymarsky:2009cm}). Following \cite{Dymarsky:2009cm}, it is possible to introduce
new coordinates similar to the ones introduced for the singular conifold \cite{Kuperstein:2008cq}:
\begin{equation}
y = \tau \cos \phi_1 \stackrel{\tau_0=0}{=} 0, \qquad z = \tau \sin \phi_1 \stackrel{\tau_0=0}{=} \pm \tau,
\end{equation}
where the $+$ sign corresponds to the $D7$-brane branch and the $-$ sign corresponds to the $\overline{D7}$-branch.
It is clear that in order to study realistic particle physics scenarios, one needs to have $\tau_0 >0$ (e.g., to separate the mass scale of the mesons from the mass scale of the glueballs). However, we have not been able to obtain a numerical solution of the coupled differential equations for the background gauge field for $\tau_0 \neq 0$. Therefore, 
we will commence our investigation of the mass spectra for (axial-) vector mesons with the antipodal embedding corresponding to $\tau_0=0$ and leave the more involved case of general $\tau_0$ as a future problem\footnote{As discussed in some detail in \cite{Dymarsky:2010ci}, the $\tau_0 \sim 0$ or $r_0 \sim r_{\epsilon}$ regime is of  great phenomenological interest, since all meson masses are of the same order $m_{\rm{gb}}$ and thus an approximate cancelation of the attractive and repulsive nuclear forces (as observed in nature) can possibly occur in this regime. A related important issue is the question whether the pseudo-Goldstone boson $\sigma$, which is parametrically lighter than the lightest vector meson for $\tau_0 >>0$, remains lighter in the $\tau_0 \sim 0$ regime, resulting in a net attractive nuclear potential at large distances.}. The general strategy is to employ a shooting method as discussed previously in, e.g., \cite{Sakai:2003wu, Bayona:2010bg}.\\
\noindent
{\bf Antipodal configuration.} In this extremal case, the functions $C_0(\tau)$ and $D_0(\tau)$ read
\begin{eqnarray}
C_0(\tau) &=& h(\tau)\frac{L_0(\tau)}{K(\tau)} \cosh \tau + \left(A_7(\tau)-b(\tau)\right) A_7'(\tau),\\
D_0(\tau) &=& \frac{K(\tau)}{L_0(\tau)} \cosh \tau,
\end{eqnarray}
where $A_7(\tau)$ is the rescaled version of the function given in (\ref{eq:A7tau0}). Moreover, $L_0(\tau)=1$ and $L_0'(\tau)=0$. We can rewrite the equations of motion (\ref{eq:eom}) as
\begin{equation}
\psi_n ''(\tau) + \frac{D_0'(\tau)}{D_0(\tau)} \psi_n ' (\tau) + \frac{C_0(\tau)}{D_0(\tau)}  \lambda_n \psi_n (\tau) =0, 
\end{equation}
where
\begin{eqnarray}\label{eq:coefffunct}
\frac{D_0'(\tau)}{D_0(\tau)} &=& \frac{3(\tau \coth \tau -1)-\sinh^2 \tau}{3(\sinh \tau \cosh \tau - \tau)} + \tanh \tau,\\
\frac{C_0(\tau)}{D_0(\tau)} &=& \frac{h(\tau) }{K^2(\tau)} + \frac{2}{3} \frac{(A_7(\tau) -b(\tau))^2}{K(\tau)^4 \cosh^2 \tau}.
\end{eqnarray}
We are now interested in finding the asymptotic behavior of the wave functions $\psi_n(\tau)$. To this end, we expand them into a Frobenius series
\begin{equation}
 \psi_n(\tau) = \pm e^{-\alpha \tau} \sum_{m=0}^{\infty} \alpha_{n,m} e^{-\gamma m \tau}.
\end{equation}
Note that
\begin{eqnarray}\label{eq:asympt}
\lim_{\tau \rightarrow \infty} \frac{D_0'(\tau)}{D_0(\tau)} &=& \frac{2}{3},\\
\lim_{\tau \rightarrow \infty} \frac{C_0(\tau)}{D_0(\tau)} &=& 0.
\end{eqnarray}
From this behavior and imposing the normalization condition (\ref{eq:norm}), we find that asymptotically the normalizable solution must have $\alpha=\frac{2}{3}$, i.e., it goes like
\begin{equation}\label{eq:psiexp}
\psi_n (\tau) = \pm e^{-\frac{2 \tau}{3}} \widetilde{\psi}_n(\tau), 
\end{equation}
where $\widetilde{\psi}_n(\tau \rightarrow \infty)=$ const. at the boundary. It is easy to check that $\widetilde{\psi}_n(\tau)$ satisfies the differential equation
\begin{equation}\label{eq:eomtilde}
\widetilde{\psi}_n''(\tau) + X(\tau) \widetilde{\psi}_n'(\tau) + ( Y(\tau) + \lambda_n Z(\tau)) \widetilde{\psi}_n(\tau)=0,
\end{equation} 
with 
\begin{eqnarray}
 X(\tau)&=& \frac{D_0'(\tau)}{D_0(\tau)} -\frac{4}{3},\\
 Y(\tau) &=& \frac{4}{9}-\frac{2}{3} \frac{D_0'(\tau)}{D_0(\tau)},\\
 Z(\tau) &=& \frac{C_0(\tau)}{D_0(\tau)}.
\end{eqnarray}
\noindent
{\bf Large $\tau$ regime.} For large $\tau$, we can expand $X(\tau), Y(\tau)$ and $Z(\tau)$ as
\begin{equation}\label{eq:XYZexp}
X(\tau) = \sum_{k=0}^{\infty} x_k (\tau) e^{-2 k \tau}, \; Y(\tau) = \sum_{k=0}^{\infty} y_k (\tau) e^{-2 k \tau}, \; Z(\tau) = e^{-\frac{2 \tau}{3}} \sum_{k=0}^{\infty} z_k (\tau) e^{-\frac{k \tau}{2}},
\end{equation}
yielding
\begin{eqnarray}
x_0(\tau) = -\frac{2}{3},\; x_1 (\tau) = -\frac{16}{3}+ \frac{8}{3} \tau,\; x_2 (\tau)= \frac{4}{3}-\frac{16}{3}\tau+\frac{32}{3}\tau^2, \ldots \\
y_0(\tau) = 0,\; y_1(\tau) = \frac{32}{9} - \frac{16}{9}\tau,\; y_2(\tau)=-\frac{8}{9} +\frac{32}{9} \tau-\frac{64}{9}\tau^2, \ldots
\end{eqnarray}
The expansion of $Z(\tau)$ is a little more involved: Note that the expansion of $\frac{h(\tau)}{K^2(\tau)}$ only contributes to the $z_{4 \mathbb{N}}$ coefficients in the expansion, while the expansion of $\frac{2}{3}\frac{(A_7 - b)^2}{K^4 \cosh^2 \tau}$ produces contributions to all the other terms as well. Then,
\begin{eqnarray}
&&z_0(\tau) = 2^{2/3}\left(\frac{15}{8} + \frac{3}{2}\tau\right),\, z_1 = 0,\, z_2 = 0, \,z_3 (\tau) = 2^{2/3}\left(-\frac{300}{169}+\frac{24}{13}\tau\right) , \nonumber \\
&&z_4 (\tau) = 2^{2/3}\left(-\frac{23661}{4000}-\frac{203}{100}\tau+\frac{31}{5} \tau^2\right),\, z_5 =0, \ldots
\end{eqnarray}
By expanding 
\begin{equation}
 \widetilde{\psi}_n(\tau) = \sum_{k=0}^{\infty} \alpha_{n,k}(\tau) e^{-\frac{k\tau}{6}}, 
\end{equation}
and using (\ref{eq:eomtilde}), one finds the following recursion relation for the coefficients $\alpha_{n,k}$:
\begin{eqnarray}\label{eq:recursion}
 \alpha_{n,k}'' - \frac{k}{3} \alpha_{n,k}'+ \frac{k^2}{36} \alpha_{n,k}
 &+& \sum_{m=0}^{\left[\frac{k}{12}\right]}\left[ x_m \left( \alpha_{n,k-12m}' -\frac{1}{6} (k-12m) \alpha_{n,k-12m}\right)+ y_m \alpha_{n,k-12m} \right] \nonumber\\
 &+& \lambda_n \sum_{m=0}^{\left[\frac{k-4}{3}\right]} z_m \alpha_{n,k-3m-4}=0.
\end{eqnarray}
Setting $ \alpha_{n,0}=1$, we obtain the first few coefficients as
\begin{eqnarray}
 && \alpha_{n,1} = \alpha_{n,2} =\alpha_{n,3} = 0,\;  \alpha_{n,4} = -\frac{27}{32}2^{2/3}\lambda_n (7+2\tau),\nonumber \\
 &&\alpha_{n,5} = \alpha_{n,6} = \alpha_{n,7}= 0, \, \alpha_{n,8} = \frac{729}{1024} 2^{1/3}\lambda_n^2 (769 + 196 \tau + 16 \tau^2),\nonumber \\
 && \alpha_{n,9} = \alpha_{n,10} = \alpha_{n,11}= 0, \, \alpha_{n,12} = -\frac{3}{8}
 +\frac{\tau}{3}+ 2^{1/3} \lambda_n^2 \left( \frac{37665}{4096} + \frac{3159}{512} \tau +\frac{243}{256} \tau^2 \right),\ldots \nonumber
\end{eqnarray}
We can now repeat the same calculation for the $\overline{D7}$-brane branch by using the SD background gauge field $\overline{A}_7 (\tau)$ instead of 
$A_7(\tau)$ and replacing the expressions for $C_0(\tau)$ and $Z(\tau)$ with the appropriate $\overline{D7}$ expressions, namely
\begin{equation}\label{eq:coefffunct2}
\overline{Z} (\tau) = \frac{\overline{C}_0}{D_0} =  \frac{h(\tau) }{K^2(\tau)} + \frac{2}{3} \frac{(\overline{A}_7(\tau) -b(\tau))^2}{K(\tau)^4 \cosh^2 \tau}.
\end{equation}
The appropriate large $\tau$ expansion for $\overline{Z}(\tau)$ turns out to be
\begin{equation}
 \overline{Z}(\tau)= e^{-\frac{2 \tau}{3}} \sum_{k=0}^{\infty} \overline{z}_k (\tau) e^{-\frac{k \tau}{6}}.
\end{equation}
Again, we can employ a similar ansatz as before, $\overline{\psi}_n = \pm e^{-\frac{2}{3}\tau}\widetilde{\overline{\psi}}_n $ with $\widetilde{\overline{\psi}}_n = \sum_k^{\infty} \overline{\alpha}_{n,k} e^{-\frac{k\tau}{6}}$ and
$\overline{\alpha}_{n,0}=1$. 
The corresponding recursion relation is given by
\begin{eqnarray}
\overline{\alpha}_{n,k}'' - \frac{k}{3} \overline{\alpha}_{n,k}'+ \frac{k^2}{36} \overline{\alpha}_{n,k}
 &+& \sum_{m=0}^{\left[\frac{k}{12}\right]}\left[ x_m \left( \overline{\alpha}_{n,k-12m}' -\frac{1}{6} (k-12m) \overline{\alpha}_{n,k-12m}\right)+ y_m \overline{\alpha}_{n,k-12m} \right] \nonumber\\
 &+& \lambda_n \sum_{m=0}^{k-4} z_m \overline{\alpha}_{n,k-m-4}=0.
\end{eqnarray}
{\bf Strategy.}
All this can now be used as input data to solve the equations of motion (\ref{eq:eom}) numerically via a shooting technique.   
Since the equations of motion for the (axial-) vector mesons depend on the background gauge field $A_7(\tau) \neq \overline{A}_7(\tau)$, the coefficient functions are different for the $D7$- and $\overline{D7}$-brane branches (cf. eqs.~(\ref{eq:coefffunct}),(\ref{eq:coefffunct2})). Therefore we do not expect the wave functions $\psi_n$ do exhibit parity symmetry about the origin, as was the case, e.g., in the Kuperstein-Sonnenschein \cite{Bayona:2010bg} and Sakai-Sugimoto \cite{Sakai:2004cn} models. However, in the large $\tau$ limit, the functions $\frac{C_0(\tau)}{D(\tau)}$ and $\frac{\overline{C}_0(\tau)}{D(\tau)}$, involving the background 
gauge field $A_7 (\tau)/\overline{A}_7(\tau)$, tend to zero. It is therefore to be expected that the parity symmetry will be restored in this limit.
As in e.g. \cite{Bayona:2010bg}, we are dealing with differential equations of the Sturm-Liouville type. 
We find it convenient to work with the $z=\pm \tau$ coordinate introduced above and denote the wave functions defined on the full domain of $z$ by $\Psi_n(z)$. Then we will treat the problem of finding a discrete spectrum of (axial-) vector mesons living on the full $D7/\overline{D7}$-brane embedding as a single differential equation on the complete domain of $z$, $z \in (-\infty , +\infty)$, as the natural extension of eq.~(\ref{eq:eom}),
\begin{equation}\label{eq:eomz}
 \Psi''_n (z) + \frac{D'_0(z)}{D_0(z)} \Psi'_n(z) + \lambda_n \frac{C_0(z)}{D_0(z)} \Psi_n(z) =0,
\end{equation}
where $D_0(z)$ is the straightforward extension of $D_0(\tau)$, and 
\begin{equation}
C_0(z) := \left\{ \begin{array}{cc} C_0(\tau) & \text{for}\; z = \tau \;\text{positive}, \\ \overline{C}_0(\tau) & \text{for}\; z = -\tau \;\text{negative}. \end{array} \right.
\end{equation}
Imposing the Frobenius expansion, eq.~(\ref{eq:psiexp}), as the starting point for the numerical evolution, which correponds to the boundary condition $\Psi_n(z= +\infty)=0$, we were able to numerically solve the equations of motion (\ref{eq:eomz}), evolving them backwards to $z \rightarrow -\infty$. 
Note that the coefficient functions $\frac{D'_0(z)}{D_0(z)}$ and $\frac{C_0(z)}{D_0(z)}$ are continuous, while their first derivatives have a discontinuity at the origin\footnote{In physical applications, a discontinuity usually signals an abrupt change in the propagation medium of a wave and can be treated be demanding appropriate ``junction conditions'' across the discontinuity.}.
In accordance with general Sturm-Liouville theory, we were able to find a discrete spectrum of (axial-) vector mesons by imposing $\Psi_n(z\rightarrow - \infty)=0$ at the opposite boundary.\\
\noindent
{\bf Numerical results.} 
There will be two distinct cases according to the large $z$ asymptotics, i.e., depending on whether the asymptotic wave functions approach zero from above or below:
\begin{itemize}
 \item Vector mesons (V): Two identical types of solutions for $\lim_{|z|\rightarrow \infty}\Psi_{2n-1}(z) = 0^+$ or $0^-$. 
 \item Axial-vector mesons (A): Two identical types of solutions corresponding to \\ $\lim_{z \rightarrow \pm \infty} \Psi_{2n}(z) = 0^{\pm}$ or $0^{\mp}$. 
\end{itemize}
The corresponding eigenvalues $\lambda_n$ are listed in table~\ref{tab:lambda}. 
Note that, in accordance with the discussion above, there are two types of solutions for small $|z|$, namely the ``cosine-like`` and the ''sine-like`` solutions and there is an approximate parity symmetry. Again, the deviations from parity invariance are due to $A_7(\tau)\neq \overline{A}_7(\tau)$.

\begin{figure}[ht] 
\begin{center}
\epsfig{file=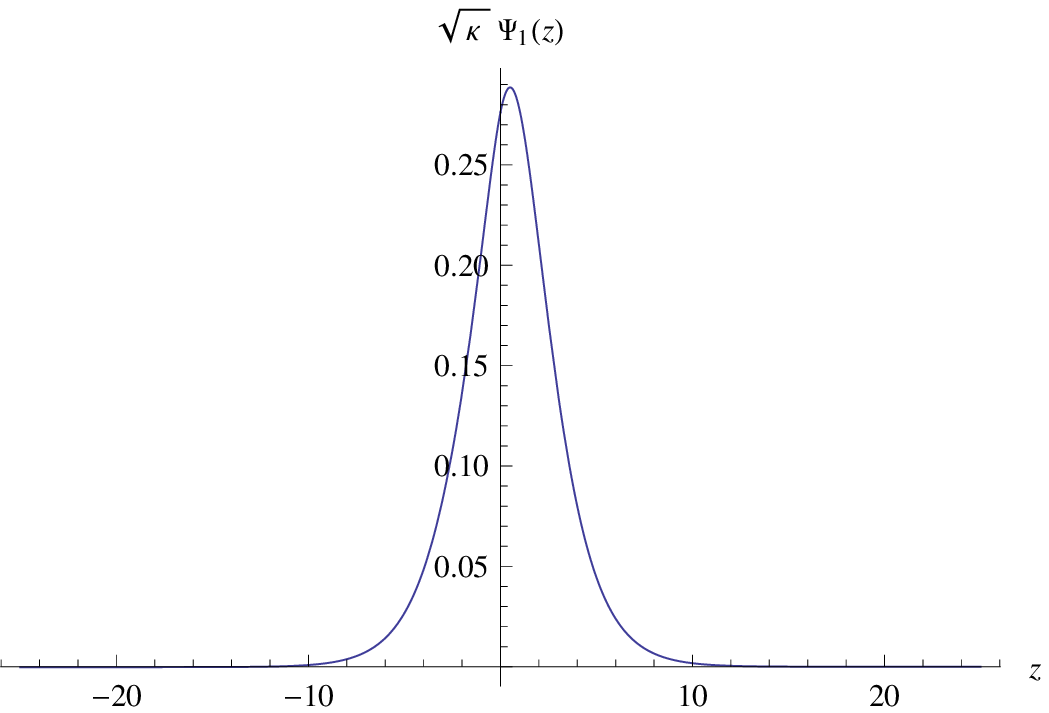, width=6cm}
\epsfig{file=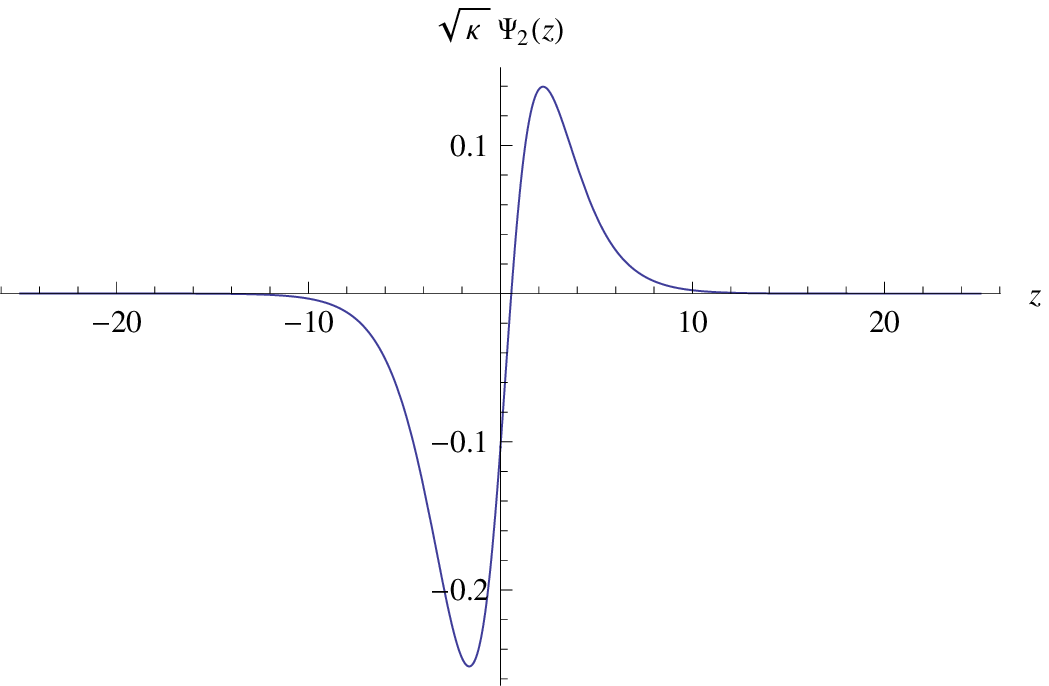, width=6cm}
\epsfig{file=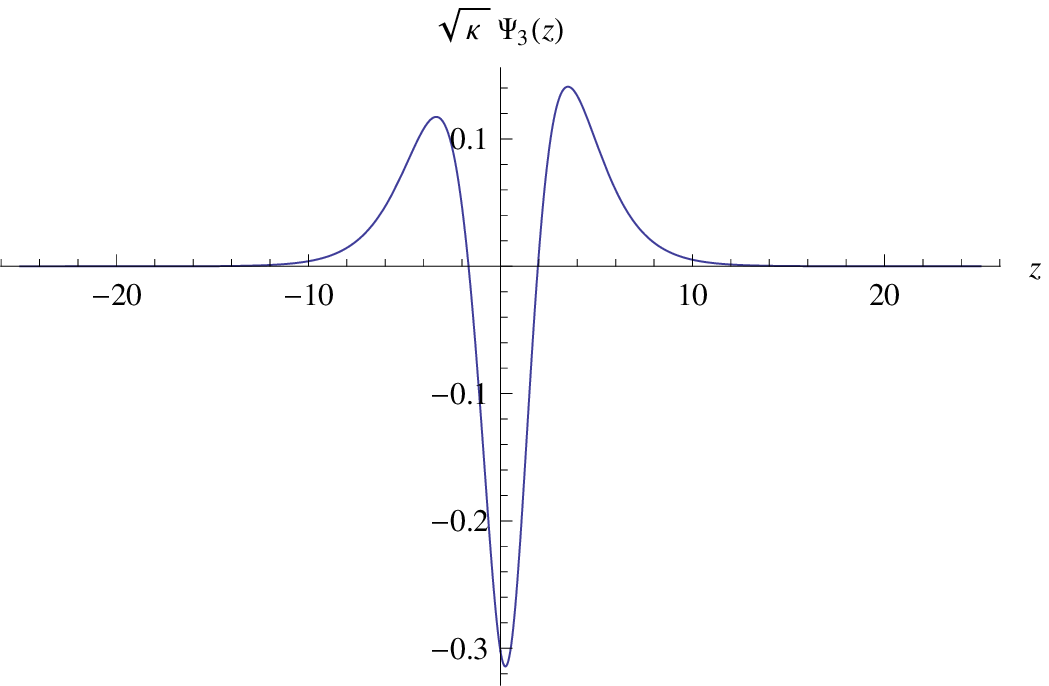, width=6cm}
\epsfig{file=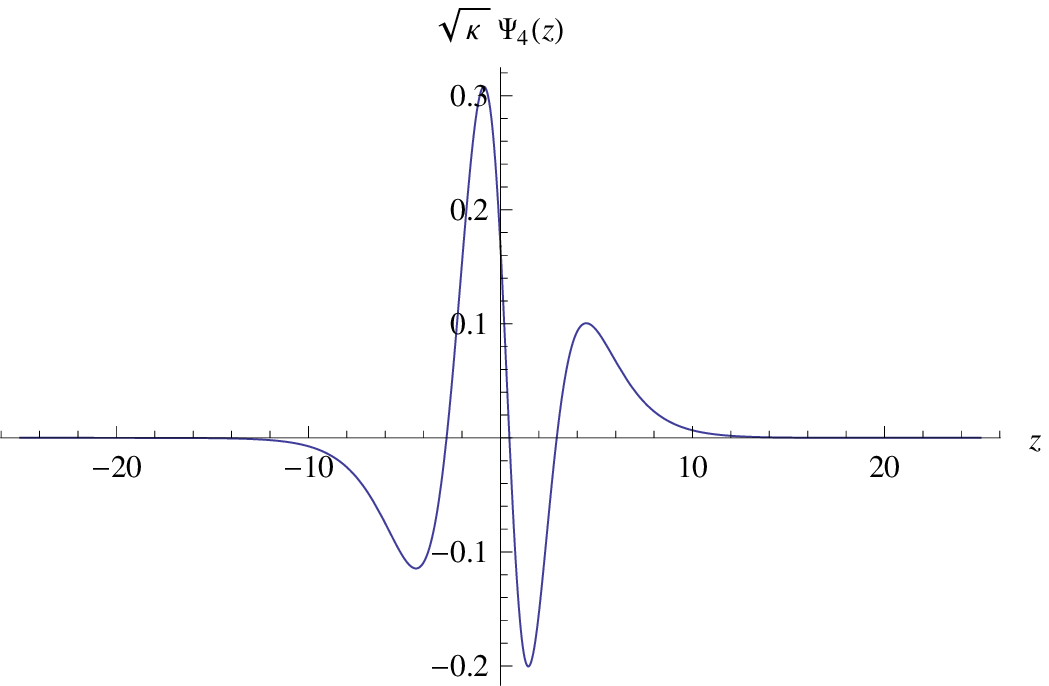, width=6cm}
\epsfig{file=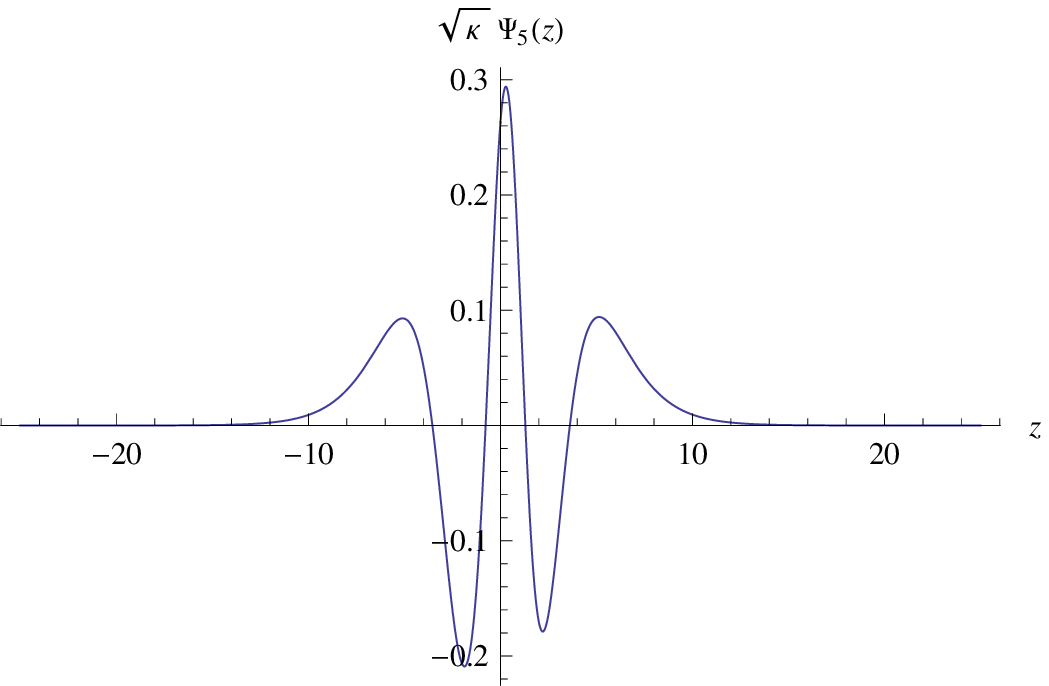, width=6cm}
\epsfig{file=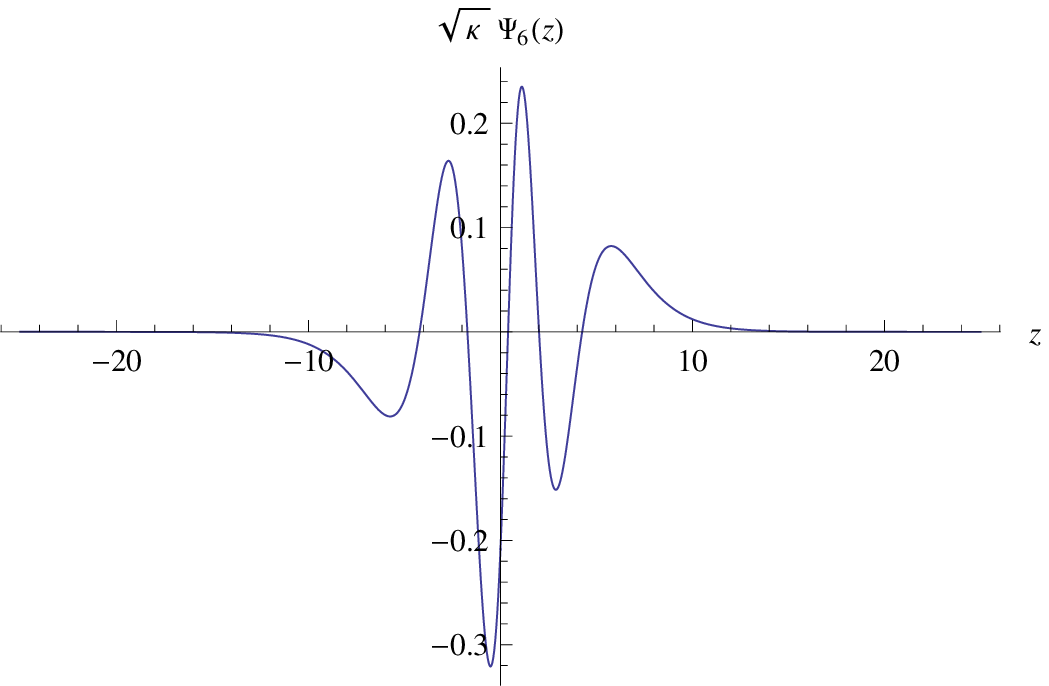, width=6cm}
\end{center}
\caption{(Normalized) wave functions $\sqrt{\kappa} \Psi_n(z)$ for the first three vector mesons $n=1,3,5$ and the first three axial-vector mesons $n=2,4,6$.}
\label{fig:wavefct}
\end{figure}

\begin{table}[h]
\begin{center}
\begin{tabular}{|c||c|c|c|c|c|c||} \hline
 $n$ & 1 & 2 & 3 & 4 & 5 & 6  \\ \hline
 $\lambda_{2n-1}=\frac{M_{2n-1}^2}{M_{\ast}^2}$ & 0.1310 & 0.4785 & 1.1081 & 2.0267 & 3.2238 & 4.6948  \\ \hline
 $\lambda_{2n}=\frac{M_{2n}^2}{M_{\ast}^2}$     & 0.2582 & 0.7785 & 1.5552 & 2.6004 & 3.9217 & 5.5214  \\ \hline

\end{tabular}
\end{center}
\caption{Some numerical values for the dimensionless (axial-) vector meson masses.}\label{tab:lambda}
\end{table}
\noindent
We found solutions $\Psi_n(z)$ and the corresponding eigenvalues $\lambda_n$ for the first 20 eigenvalues.
The results for a selection of wave functions are shown in figure~\ref{fig:wavefct}. 

\begin{figure}[ht] 
\begin{center}
\epsfig{file=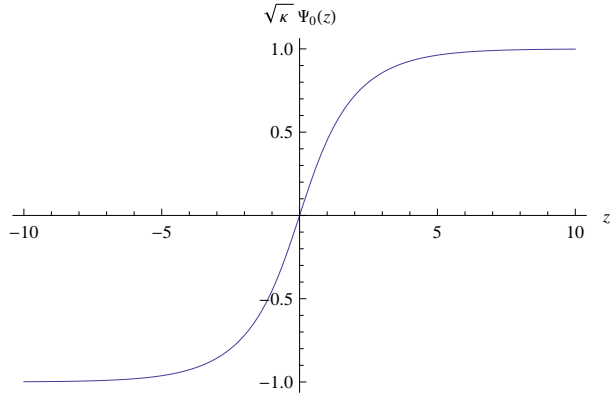, width=8cm}
\end{center}
\caption{Zero mode $\sqrt{\kappa} \Psi_0(z)$.}
\label{fig:zeromode}
\end{figure}

\noindent There is also a non-renormalizable zero mode subject to $\partial_z \Psi_0(z) \propto D(z)^{-1}$ and \\
$\lim_{z\rightarrow \pm \infty} \Psi_0(z)=\pm 1$, associated with the pions in the theory (see discussion in \cite{Bayona:2010bg}, sect. 4). This mode is shown in figure \ref{fig:zeromode}.

\noindent It is possible to fix the mass scale $M_{\ast}$ by identifying the lightest vector meson $v^{(1)}$ with the $\rho$-meson $\rho(770)$ of mass $M_{\rho} = 775 \text{MeV}$, yielding $M_{\ast}^2= \frac{M_{\rho}^2}{\lambda_1} = (2141 \text{MeV})^2$. Similarly, the lightest axial vector meson $a^{(1)}$ can be identified with $a_1 (1260)$, the second lightest vector meson $v^{(2)}$ with $\rho (1450)$, and so forth. Moreover, it would be possible to fix $\kappa$ so as to obtain a realistic value for the pion decay constant. In table \ref{tab:ratio}, we compare the mass ratios of the first few mesons in our model (DKS) (for the antipodal embedding $\tau_0=0$) with the corresponding ratios found in the Kuperstein-Sonnenschein (KS) \cite{Bayona:2010bg} and Sakai-Sugimoto (SS) models \cite{Sakai:2005yt} and experiments \cite{Nakamura:2010zzi}.

\begin{table}[ht]
\begin{center}
\begin{tabular}{|c||c|c|c|c|}
\hline
k & $\left(\frac{\lambda_{k+1}}{\lambda_1}\right)_{\text{DKS}}$ & $\left(\frac{\lambda_{k+1}}{\lambda_1}\right)_{\text{KS}}$  & $\left(\frac{\lambda_{k+1}}{\lambda_1}\right)_{\text{SS}}$ & $\left(\frac{\lambda_{k+1}}{\lambda_1}\right)_{\text{exp.}}$ \\\hline
1 & 1.97 & 2.68 & 2.34 & $\sim$ 2.51 \\\hline
2 & 3.65 & 5.36 & 4.29 & $\sim$ 3.56 \\\hline
3 & 5.94 & 8.88 & 6.79 & $\sim$ 4.54  \\\hline
4 & 8.46 & 13.3 & 9.85  & $\sim$ 4.15  \\\hline
\end{tabular}
\end{center}
\caption{Ratios $M_{k+1}^2/M_1^2 = \lambda_{k+1}/\lambda_1$ for $k=1,\ldots, 5$. The numerical values in the DKS case are obtained for the antipodal configuration $\tau_0=0$. One should note that the DKS ratios are closer to the experimental values than other holographic models, while for $k>3$ neither model matches the experiments well.}\label{tab:ratio}
\end{table}
\newpage
\subsection{Scalar mesons}
In this section, we are interested in studying the scalar fluctuations around the flavor brane embeddings.
 As for the vector mesons, we will derive a five-dimensional effective action up to quadratic order for the 5d scalar fluctuation modes corresponding to (pseudo-) scalar mesons.
 The potentially relevant modes stem from the Kaluza-Klein reduction of the 8d scalars fluctuations $\delta \theta_1$,  $\delta \phi_1$ transverse to 
the $D7$-brane embedding and scalar modes of the 8d gauge field, i.e., $\delta A_5, \delta A_6, \delta A_7$, respectively. \\
It is important to note\footnote{We would like to thank A. Dymarsky for clarifying this point to us.} that, for the 
antipodal configuration $\tau_0=0$, there is an unbroken $U(1)$ that rotates the $S^2$ around the two antipodal points, but leaves the $D7$-branes invariant. 
In this extremal case there is also an emergent $U(1)_R$ symmetry that shifts the angle $\psi$ by a constant, acting on the $D7$-branes but leaving the induced metric invariant.
 While the unbroken $U(1)$ symmetry may rotate fluctuations of $\theta_1$ into fluctuations of $\phi_1$, we will see that from the expansion of the DBI-WZ action 
there are no mixing terms between the $\delta \theta_1$ and $\delta \phi_1$ fluctuations and, interestingly , the scalar fluctuation  $\delta A_7$. 
This happens accidently only for the antipodal case, where $\phi_1'(\tau)=0$ classically.

The relevant fluctuations for a QCD-like theory are singlet states\footnote{Non-singlet states with nontrivial charges under the $SU(2)\times U(1)_R$ isometry 
have no counterpart in the dual QCD and will therefore be disregarded.} under the $SU(2)\times U(1)_R$ isometry of the internal $S^3$. 
The simplest way to preserve  the $SU(2)\times U(1)_R$ isometry discussed above would be to disregard 
$\delta \theta_1, \delta \phi_1, \delta A_5$ and $\delta A_6$ fluctuations, which would automatically break the isometry symmetry. It is possible that there exists a nontrivial singlet state, i.e., some linear combination of all the fluctuations. This will be left to a future investigation.  
In this subsection, we calculate the effective actions for all the scalar fluctuations. We will find non-trivial mixing terms 
between the $\delta \phi_1$ ($\delta \theta_1$) and the $\delta A_5$ ($\delta A_6$) fluctuations\footnote{These terms arise from the metric components describing
 the non-trivial fibration of the $S^3$ over the $S^2$ which were absent in the original treatment of Sakai and Sonnenschein \cite{Sakai:2003wu}.}. 
We will see that the  $\delta A_7$ fluctuation does not mix with the other fluctuations suggesting its interpretation as a singlet and
 therefore a good candidate for a scalar meson in the dual theory. However, after performing a Kaluza-Klein expansion we do not find
a consistent set of orthonormal eigenfunctions due to the discontinuity at the origin of one of the Sturm-Liouville coefficients arising from the DBI-WZ expansion. However, it is quite clear that this discontinuity is not a real physical effect (e.g., there is no charge located at the origin), but rather an artifact of our choice of coordinates. Therefore a smooth and stable spectrum of scalar mesons should exist. Unfortunately, we have not yet been able to numerically solve the corresponding Sturm-Liouville problem in a completely satisfactory way. 

\subsubsection{Effective action for scalars fluctuations}

Consider the coordinate fluctuations 
\beqa
\delta \theta_1 (x , \tau) &=& ( 2 \pi \alpha') \hat \theta_1 (x , \tau)  \, , \cr 
\delta \phi_1 (x , \tau)  &=& ( 2 \pi \alpha') \hat \phi_1 (x , \tau) \, .
\eeqa
As a consequence, the induced metric now contains linear and quadratic terms in the fluctuations
\beqa
&& \delta (g_{(8)})_{\mu \nu} =  G_{\theta_1 \theta_1} \partial_\mu \delta \theta_1 \partial_\nu \delta \theta_1 + G_{\phi_1 \phi_1} \partial_\mu \delta \phi_1 \partial_\nu \delta \phi_1  \, , \cr
&& \delta (g_{(8)})_{\mu \tau} = G_{\theta_1 \theta_1} \partial_\mu \delta \theta_1 \partial_\tau \delta \theta_1 + G_{\phi_1 \phi_1} \partial_\mu \delta \phi_1  \partial_\tau\delta \phi_1 \, \cr
&& \delta (g_{(8)})_{\mu i} =  G_{\theta_1 1} \partial_\mu \delta \theta_1  \delta_{i 1} + G_{\phi_1 2} \partial_\mu \delta \phi_1 \delta_{i 2} + \delta G_{\phi_1 3} \partial_\mu \delta \phi_1 \delta_{i 3} \, , \cr
&& \delta (g_{(8)})_{\tau \tau} =  G_{\theta_1 \theta_1} (\partial_\tau \delta \theta_1)^2  + G_{\phi_1 \phi_1}  (\partial_\tau \delta \phi_1)^2 \cr 
&& \delta (g_{(8)})_{\tau i} =G_{\theta_1 1} \partial_\tau \delta \theta_1  \delta_{i 1} + G_{\phi_1 2}  \partial_\tau \delta \phi_1 \delta_{i 2} + \delta G_{\phi_1 3}  \partial_\tau \delta \phi_1 \delta_{i 3}\, ,
\eeqa
where $G_{PQ}$ is the 10-d metric of the deformed conifold (see appendix B). 
On the other hand, the gauge field fluctuations
\beqa
\delta A_5 (x , \tau) &=& (2 \pi \alpha') a_5 (x , \tau) \, , \cr 
\delta A_6 (x , \tau) &=& (2 \pi \alpha') a_6 (x , \tau) \, , \cr
\delta A_7 (x , \tau) &=& (2 \pi \alpha') a_7 (x , \tau) \, ,
\eeqa
induce the field strength fluctuation
\beqa
\delta {\cal F}_{\tau 1} &=& -  \frac{b \partial_\tau \delta \phi_1 }{\cosh \tau}  + \partial_\tau \delta A_5  \quad , \quad  \delta {\cal F}_{\tau 2} = \partial_\tau \delta A_6 \quad , \quad \delta {\cal F}_{\tau 3}  = \partial_\tau \delta A_7 \cr 
\delta {\cal F}_{23} &=& - \delta A_5  \quad , \quad  \delta {\cal F}_{13} =  \delta A_6 \quad, \quad \delta {\cal F}_{12} =  - \delta A_7 \, , \cr 
\delta {\cal F}_{\mu 1} &=& \partial_\mu \delta A_5 \quad , \quad \delta {\cal F}_{\mu 2} = \partial_\mu \delta A_6 \quad , \quad \delta {\cal F}_{\mu 3} = \partial_\mu \delta A_7 \, .   
\eeqa

From the linear and quadratic terms in the DBI expansion we extract the actions for the scalar fluctuations 
\beqa
S_{DBI}^{(\delta \phi_1 , \delta A_5)}   &=&  - \frac{\mu_7}{2 g_s} \int d^4 x d \tau d^3 \Omega \sqrt{|E|}  
\Big \{  
 h^{1/2} \eta^{\mu \nu}  
\Big [(G_{\phi_1 \phi_1} - E^{22} G_{\phi_1 2}^2 ) \partial_\mu \delta \phi_1 \partial_\nu \delta \phi_1 \cr 
&+& E^{11} \partial_\mu \delta A_5 \partial_\nu \delta A_5 - 2 E^{12} G_{\phi_1 2} \partial_\mu \delta \phi_1 \partial_\nu \delta A_5
\Big ] \cr
&+& E^{\tau \tau}  \Big [
(G_{\phi_1 \phi_1} - E^{22} G_{\phi_1 2}^2 ) (\partial_\tau \delta \phi_1)^2 
+ E^{11} \left (  - \frac{b \partial_\tau \delta \phi_1}{\cosh \tau} + \partial_\tau \delta A_5 \right )^2 \cr
&-& 2 E^{12} G_{\phi_1 2} \partial_\tau \delta \phi_1 \left (  - \frac{b \partial_\tau \delta \phi_1}{\cosh \tau} + \partial_\tau \delta A_5 \right ) \Big ] \cr 
&-& 2 E^{\tau 3} \left [  E^{12} \delta A_5 \left (  - \frac{b \partial_\tau \delta \phi_1}{\cosh \tau} + \partial_\tau \delta A_5 \right ) + E^{22} G_{\phi_1 2} \delta A_5 \partial_\tau \delta \phi_1  \right ] \cr 
&+& E^{22} E^{33} (\delta A_5)^2 
\Big \} \, ,
\eeqa
\beqa 
S_{DBI}^{(\delta \theta_1 , \delta A_6)}  &=& - \frac{\mu_7}{2 g_s} \int d^4 x d \tau d^3 \Omega \sqrt{|E|} 
\Big \{  h^{1/2} \eta^{\mu \nu} \Big [ (G_{\theta_1 \theta_1} - E^{11} G_{\theta_1 1}^2 ) \partial_\mu \delta \theta_1 \partial_\nu \delta \theta_1  \cr 
&+&  E^{22} \partial_\mu \delta A_6 \partial_\nu \delta A_6 + 2 E^{12} G_{\theta_1 1} \partial_\mu \delta \theta_1 \partial_\nu \delta A_6 \Big ] \cr
&+&  E^{\tau \tau}  \Big [ (G_{\theta_1 \theta_1} - E^{11} G_{\theta_1 1}^2 ) (\partial_\tau \delta \theta_1)^2  + E^{22} (\partial_\tau \delta A_6)^2 
+ 2 E^{12} G_{\theta_1 1} \partial_\tau \delta A_6 \partial_\tau \delta \theta_1 \Big ]  \cr 
&-& 2 E^{\tau 3} \left [ E^{12} \delta A_6 \partial_\tau \delta A_6 - E^{11} G_{\theta_1 1} \delta A_6 \partial_\tau \delta \theta_1 \right ] \cr 
&+& E^{11} E^{33} (\delta A_6)^2  \Big \} \,,
\eeqa

\beqa
S_{DBI}^{(\delta A_7)} &=& - \frac{\mu_7}{2 g_s} \int d^4 x d \tau d^3 \Omega \sqrt{|E|} 
\Big \{ - 2 E^{\tau 3} \partial_\tau \delta A_7 + 2 E^{12} \delta A_7   \cr 
&+& E^{33} h^{1/2} \, \eta^{\mu \nu} \partial_\mu \delta A_7 \partial_\nu \delta A_7 +  E^{\tau \tau} E^{33} (\partial_\tau \delta A_7)^2  \cr 
&+& E^{11} E^{22} (\delta A_7)^2 - 2  E^{12} E^{\tau 3} (\partial_\tau \delta A_7) \delta A_7 \Big \}  \,.
\eeqa
The tensor $E^{MN}$ is defined as the inverse of $E_{MN}=(g_{(8)})_{MN} + {\cal F}_{MN}$ (see appendix B for details).
The WZ action is given by 
\beqa
S_{WZ} &=& - \frac{\mu_7}{2} \int  {\cal F} \wedge {\cal F} \wedge C_{(4)} \cr 
&=& - \frac{\mu_7}{8 g_s} \epsilon_{\alpha \beta \rho \sigma} \int h^{-1} {\cal F}_{\alpha \beta} {\cal F}_{\rho \sigma} d^4 x d \tau  d \Omega^1  d^3 \Omega
\eeqa
where $\alpha=(\tau,\Omega^i)$. This action contributes to the scalar fluctuations as 
\beqa
S_{WZ}^{(\delta A_5 , \delta A_6 , \delta A_7 )} &=&  - \frac{\mu_7}{ g_s} \int  d^4 x d \tau  d^3 \Omega \, h^{-1}
\Big \{  - {\cal F}_{\tau 3}  \delta A_7 + {\cal F}_{12}  \partial_\tau \delta A_7   \cr 
&-&  \delta A_5 \left ( - \frac{ b \partial_\tau \delta \phi_1}{\cosh \tau} + \partial_\tau \delta A_5  \right ) 
- (\partial_\tau \delta A_6) \delta A_6 - (\partial_\tau \delta A_7) \delta A_7  \Big \}  \, , 
 \eeqa

As expected,  the linear term arising from the WZ action cancels with the linear term arising from the DBI action. The results for the quadratic terms 
show that there is a non-trivial mixing between the scalars $\delta \phi_1$ ($\delta \theta_1$) and $\delta A_5$ ($\delta A_6$). A linear combination 
of these fluctuations is not enough to diagonalize the action. A simple solution to this problem is to set $\delta \phi_1 = \delta \theta_1 = \delta A_5 = \delta A_6 = 0$
but we don't discard the possibility of a non-trivial singlet combination. 

\subsubsection{Scalar fluctuations of $A_7$}
The only remaining scalar mode of interest is the one associated with fluctuations of $A_7$. Summing the DBI and WZ terms we can extract the action for the fluctuation $a_7$  
\begin{eqnarray}
S^{A_7}_{5d, \text{eff.}}&=& -\overline{\kappa} \int \d x^4 \d \tau \left( \frac{1}{2} G(\tau)\, \eta^{\mu \nu}\, \partial_{\mu} a_7 (x^{\mu},\tau)\, \partial_{\nu} a_7 (x^{\mu},\tau) + M_{\ast}^2 H(\tau)\, \partial_{\tau} a_7 (x^{\mu},\tau)\, \partial_{\tau} a_7 (x^{\mu},\tau)\right. \nonumber \\
&& \qquad \qquad \qquad \left. + M_{\ast}^2 I(\tau)\, \left(a_7 (x^{\mu},\tau)\right)^2 +  M_{\ast}^2 J(\tau)\, \left(\partial_{\tau}a_7 (x^{\mu},\tau)\right) a_7 (x^{\mu},\tau) \right),
\end{eqnarray}
where $\overline{\kappa} = \frac{1}{2} \mu_7 g_s^{-1} (\pi \alpha)^2 \text{Vol}\,S^3 \eps^{4/3}$, $ M_{\ast}^2=6 \frac{\eps^{4/3}}{(g_s M \alpha')^2}$ as before and we have again utilized the rescaled dimensionless functions introduced in (\ref{eq:dimless}) to define
\begin{eqnarray*}
G(\tau) &=& 2 K(\tau) \cosh \tau = 2 \hat D(\tau),\\
H(\tau) &=& \left(\frac{\hat{h}(\tau) K(\tau)}{\cosh \tau}+ \frac{2}{3} \frac{\left(\hat{A}_7 - \hat{b}\right)^2}{K(\tau) \cosh^3 \tau} \right)^{-1} K^4 (\tau) = \frac{\hat D^2(\tau)}{\hat C(\tau)}, \\
I(\tau) &=& \left ( \frac{4 }{9 K^6 (\tau) \cosh^2 \tau} \right )  H(\tau) , \\
J(\tau) &=& \frac{8}{9}\left(\frac{\hat{h}(\tau) K(\tau)}{\cosh \tau}+ \frac{2}{3}  \frac{\left(\hat{A}_7 - \hat{b}\right)^2}{K(\tau) \cosh^3 \tau} \right)^{-1} \left( \frac{\left(\hat{A}_7 - \hat{b}\right)^2}{\hat{h}(\tau)K(\tau)\cosh^3 \tau}\right)- \frac{4}{3} \hat{h}^{-1}(\tau),\\
&=& - \left ( \frac{4}{3 K^3(\tau) \cosh \tau } \right ) H(\tau) ,
\end{eqnarray*}
where the second term in the first line of $J(\tau)$ is a contribution from the WZ term .
We will again drop the hats and superscripts for notational convenience and continue to use barred quantities $\overline{H}(\tau), \overline{I}(\tau)$ and $\overline{J}(\tau)$ to denote functions defined on the left (anti-D-brane) branch, i.e., with the appropriate choice of $\hat{A}_7$ substituted into the functions defined above.\\
\noindent
{\bf Strategy.}
The five-dimensional fluctuations on the two branches can then be decomposed as
\begin{equation*}
 a_7 (x^{\mu}, \tau) = \sum_{k=0}^{\infty} a_7^{(n)}(x^{\mu}) \phi_n(\tau).
\end{equation*}
The wave functions $\phi_n(\tau)$ have to satisfy the following equations,
\begin{eqnarray}
 \overline{\kappa} \int \d \tau G(\tau) \phi_n(\tau) \phi_m (\tau) &=&\delta_{mn}, \\
(G(\tau))^{-1} \left[- \partial_{\tau} \left( H(\tau) \partial_{\tau} \phi_n(\tau) \right)+ I(\tau) \phi_n(\tau) + J(\tau) \partial_{\tau}\phi_n(\tau)\right] &=& \frac{1}{2} \overline{ \lambda}_{n} \phi_n(\tau).
 \end{eqnarray}
implying
\begin{equation}
 \overline{\kappa} \int \d \tau \left( H(\tau) \partial_{\tau}\phi_n(\tau)\, \partial_{\tau} \phi_m (\tau) + I(\tau) \phi_n(\tau) \phi_m(\tau)+ J(\tau) \left( \partial_{\tau}\phi_n(\tau)\right)\phi_m(\tau)\right)= \frac{1}{2} \overline{\lambda}_{n} \delta_{mn}.
\end{equation}
This will lead to a four-dimensional effective Lagrangian
\begin{equation}
\frac{1}{2} \int \d^4 x \sum_n \left[ \partial_{\mu} a_7^{(n)} (x^{\mu}) \partial^{\mu} a_7^{(n)} (x^{\mu}) + \overline{\lambda}_n M_{\ast}^2 (a_7 (x^{\mu}))^2 \right].
\end{equation}
Therefore, $a_7^{(n)}$ represents a scalar meson with mass given by $\frac{1}{2}\overline{\lambda}_n M_{\ast}^2$.\\
Again, we will study the $D7$-brane branch first and then generalize to the $\overline{D7}$-brane branch. The equation of motion can be rewritten as
\begin{equation}\label{eq:SLscalar}
 \phi_n''(\tau) + \left(\frac{H'(\tau)}{H(\tau)} -\frac{J(\tau)}{H(\tau)}\right) \phi_n'(\tau) + \left( \frac{1}{2}\overline{\lambda}_n - \frac{I(\tau)}{G(\tau)} \right) \frac{G(\tau)}{H(\tau)} \phi_n = 0 \,.
\end{equation}
The leading large $\tau$ behavior of the various coefficient functions is given by
\begin{equation}
 \lim_{\tau \rightarrow \infty} \frac{H'(\tau)}{H(\tau)} = \frac{4}{3},\;
 \lim_{\tau \rightarrow \infty} \frac{G(\tau)}{H(\tau)} = 0,\;
 \lim_{\tau \rightarrow \infty} \frac{I(\tau)}{H(\tau)} = \frac{4}{9},\;
 \lim_{\tau \rightarrow \infty} \frac{J(\tau)}{H(\tau)} =-\frac{4}{3}.
\end{equation}
Therefore, it is easy to see that the normalizable solutions behave as
\begin{equation}\label{eq:eomtildephi}
\phi_n(\tau) = \pm e^{-\frac{4+2\sqrt{5}}{3}\tau} \widetilde{\phi}_n(\tau), \quad \widetilde{\phi}_n(\tau \rightarrow \infty) \rightarrow \text{const.}
\end{equation}
The resulting differential equation for $\widetilde{\phi}_n(\tau)$ reads
\begin{equation}
 \widetilde{\phi}_n''(\tau) + S(\tau) \widetilde{\phi}_n'(\tau) + (T(\tau) + \frac{1}{2}\overline{\lambda}_n U(\tau)) \widetilde{\phi}_n(\tau)=0,
\end{equation}
where
\begin{eqnarray}
S(\tau) &=& \frac{H'(\tau)}{H(\tau)}-\frac{J(\tau)}{H(\tau)} - \frac{8+ 4\sqrt{5}}{3},\\
T(\tau) &=& \frac{(4+2\sqrt{5})^2}{9}- \frac{4+2\sqrt{5}}{3} \left( \frac{H'(\tau)}{H(\tau)}-\frac{J(\tau)}{H(\tau)}\right) - \frac{I(\tau)}{H(\tau)}, \\
U(\tau) &=& \frac{G(\tau)}{H(\tau)}= 2 Z(\tau).
\end{eqnarray}
\noindent
{\bf Large $\tau$ regime.} It should be noted that at leading order in large $\tau$, the contribution of the background gauge field vanishes, since
\begin{equation}
\lim_{\tau \rightarrow \infty} (A_7(\tau) -b(\tau))^2 = \left(\frac{3}{2}\right)^3, \; \lim_{\tau \rightarrow \infty} \partial_{\tau} (A_7(\tau) -b(\tau)) = 0.  
\end{equation}
The expansion of the relevant coefficient functions reads
We can expand the coefficient functions $S(\tau), T(\tau)$ and $U(\tau)$ for large $\tau$ as
\begin{equation}
S(\tau) = \sum_{k=0}^{\infty} s_k(\tau) e^{- 2k \tau}, \; T(\tau) = \sum_{k=0}^{\infty} t_k(\tau) e^{- 2k \tau}, \; 
U(\tau) = e^{- \frac{2}{3} \tau}\sum_{k=0}^{\infty} u_k(\tau) e^{- 2k \tau}, 
\end{equation}
where the first few expansion coefficients are determined to be 
\begin{eqnarray*}
 &&s_0 (\tau)= -\frac{4\sqrt{5}}{3}-\frac{4}{5+4\tau}, \\ 
 &&s_1 (\tau)= \left(16 + \frac{8}{5}\right) \tau , \ldots \\
 &&t_0 (\tau)= \frac{4+2\sqrt{5}}{3}\frac{4}{5+4\tau}, \\ 
 &&t_1 (\tau)=\frac{4+2\sqrt{5}}{3}\left( 16(1-\tau)+ \frac{8}{5} \tau \right) +\frac{32}{9}(1-\tau), \ldots \\
 &&u_0(\tau) = 2^{5/3}\left( \frac{15}{8} + \frac{3}{2}\tau \right), \\ 
 &&u_1 (\tau)= 2^{5/3}\left[ \left( - \frac{9}{16} - \frac{9}{2}\tau \right) + \frac{1}{250}\left( -96 +555\tau + 300 \tau^2 \right) \right] + 2^{8/3} \left(  \frac{15}{8} + \frac{3}{2}\tau \right) \left( 1 -2 \tau \right), \ldots
\end{eqnarray*}
Here we included at least the leading order in $\tau$.
Expanding 
\begin{equation}
 \widetilde{\phi}_n(\tau) = \sum_{k=0}^{\infty} \beta_{n,k}(\tau) e^{-\frac{2k\tau}{3}}, 
\end{equation}
and using (\ref{eq:eomtildephi}), it is easy to find a recursion relation for $\beta_{n,k}$,
\begin{eqnarray}\label{eq:phirecursion}
 \beta_{n,k}'' - \frac{4k}{3} \beta_{n,k}'+ \frac{4k^2}{9} \beta_{n,k}
 &+& \sum_{m=0}^{k}\left[ s_m \left( \beta_{n,k-3m}' -\frac{2}{3} (k-3m) \beta_{n,k-3m}\right)+ t_m \beta_{n,k-3m} \right] \nonumber\\
 &+& \frac{1}{2}\overline{\lambda}_{n} \sum_{m=0}^{\left[\frac{k-1}{3}\right]} u_m \beta_{n,k-3m-1}=0.
\end{eqnarray}
Setting $ \beta_{n,0}=1$, we find the first non-trivial coefficient to be
\begin{equation}
 \beta_{n,1} = 27 \; 2^{-13/3} \overline{\lambda}_{n} \, \tau \left( 71- 31 \sqrt{5} + 2 (-2 +\sqrt{5}) \tau \right),\, \ldots
\end{equation}
\noindent
{\bf Numerical results.} In principle, one could proceed as before and try to solve the Sturm-Liouville problem via a shooting method. 
However, we did not obtain a stable numerical solution for the eigensystem of eq.~(\ref{eq:SLscalar}). 
This is due to the discontinuity of the coefficient $S(\tau)$ at the origin $\tau = |z| =0$ (related to our choice of coordinates). 
We tried to reformulate the problem slightly, using two different eigenvalues for the two branches and glueing the solutions smoothly at the origin.
However, in that case the set of solutions do not satisfy orthonormality. Therefore we were not able to find a stable spectrum of scalar meson fluctuations for $a_7$ on the full domain 
of the flavor $D7$-$\overline{D7}$ branes. 

\subsubsection{Scalar fluctuations of $\theta_1$ (and $\phi_1$) when $\delta A_5 = \delta A_6 = 0$.}

Finally,  we would like to analyze the effective action for the scalar fluctuation $\theta_1$ in the special case when the gauge field fluctuations are turned off, i.e.  $\delta A_5 = 0$ 
and $\delta A_6 = 0$, which is consistent with the (A)SD eqs.~(\ref{eq:ASD}) only for the case $\tau_0=0$. For this case, 
the effective action associated with fluctuations around $\theta_1^{(0)}= \frac{\pi}{2}$ takes the form 
\beqa
S^{\delta \theta_1 }_{5d, \text{eff.}} = - \frac{\tilde \kappa}{2} \int d^4 x d \tau  \left \{  A (\tau) \eta^{\mu \nu} \partial_\mu \delta \theta_1 \partial_\nu \delta \theta_1 
\,+\,   M_{\ast}^2 B(\tau) (\partial_\tau \delta \theta_1 )^2 \right \} \, ,
\eeqa
where
\beqa
\tilde \kappa &=& \frac{\mu_7}{96 g_s} \text{Vol}(S^3) \epsilon^{4/3}  (g_s M \alpha')^2 \, ,\cr 
A(\tau) &=& \hat h (\tau) \cosh^2 \tau + \frac{2}{3 K^2(\tau)} (\hat A_7 - \hat b)^2 - 4 \hat h(\tau) = \hat C(\tau) \hat D(\tau) - 4 \hat h(\tau)\, ,\cr 
B(\tau) &=& K^2(\tau) \cosh^2 \tau \left [ 1 - 4 \hat h(\tau) \left (  \hat h (\tau) \cosh^2 \tau + \frac{2}{3 K^2(\tau)} (\hat A_7 - \hat b)^2 \right )^{-1} \right ] \cr
&=& \hat D^2(\tau) - 4 \hat h(\tau) \frac{ \hat D(\tau)}{ \hat C(\tau)}  \,,
\eeqa
on the $D7$-brane. On the $\overline{D7}$-brane branch, one has to replace $A_7 (\tau)$ by $\overline{A}_7(\tau)$. 
We have made use of the same dimensionless functions defined in (\ref{eq:dimless}).The effective action for the $\delta \phi_1$ fluctuations yields a similar result.  
The five dimensional fluctuation $\theta_1(x^{\mu}, \tau)$ can be decomposed as
\begin{equation}
 \delta \theta_1(x^{\mu}, \tau) = \sum_{n=1}^{\infty} \theta_1^{(n)} (x^{\mu}) \xi_n (\tau) \, ,
\end{equation}
so that we get a four dimensional effective action 
\beqa
S^{\theta_1^{(n)} }_{4d, \text{eff.}} = - \frac12 \sum_{n=1}^{\infty} \int d^4 x \left [ \eta^{\mu \nu} \partial_\mu \theta_1^{(n)} \partial_\nu \theta_1^{(n)} 
+ M_{\ast}^2 \lambda_n (\theta_1^{(n)})^2 \right ] \, ,
\eeqa
where the normalization conditions and equations of motion for the wave functions $\xi_n(\tau)$ read
\begin{eqnarray}
\widetilde{\kappa} \int d\tau \; A(\tau) \xi_m(\tau) \xi_n(\tau) &=& \delta_{m,n}, \label{eq:normscalar} \\
- (A(\tau))^{-1} \partial_{\tau} (B(\tau) \partial_{\tau} \xi_n (\tau)) &=& \widetilde{\lambda}_n \xi_n(\tau),\label{eq:eomscalar}
\end{eqnarray}
with $\widetilde{\lambda}_n := M_n^2/M_{\ast}^2$ denoting the eigenvalues of the scalar fluctuations.
Again, we are solving a Sturm-Liouville type problem and in principle we can follow the same strategy as for the vector mesons and use a shooting method to solve the equations of motion, rewritten as
\begin{equation}
 \xi_n''(\tau) + \frac{B'(\tau)}{B(\tau)}  \xi_n'(\tau) + \frac{A(\tau)}{B(\tau)} \widetilde{\lambda}_n \xi_n(\tau) =0, \label{eqxin}
\end{equation}
where
\begin{subequations}\label{eq:scalarasympt}
\begin{eqnarray}
\frac{B'(\tau)}{B (\tau)}&=& \frac{2}{ \hat C(\tau) \hat D(\tau) - 4 \hat h(\tau) } 
\left [ \hat C(\tau) \hat D'(\tau) - 2 \hat h(\tau) \left ( \frac{\hat h'(\tau)}{\hat h(\tau)} + \frac{\hat D'(\tau)}{\hat D(\tau)} - \frac{\hat C'(\tau)}{\hat C(\tau)}\right )\right ],\\
\frac{A(\tau)}{B (\tau)}&=&  \frac{\hat C(\tau)}{\hat D (\tau)} =  \frac{\hat h(\tau) }{K^2(\tau)} + \frac{2}{3} \frac{(\hat A_7(\tau) - \hat b(\tau))^2}{K^4(\tau) \cosh^2 \tau}   ,
\end{eqnarray}
\end{subequations}
on the $D7$-brane branch. Again, on the $\overline{D7}$-brane branch, one has to replace $A_7 (\tau)$ by $\overline{A}_7(\tau)$. 

Thus we again find a discontinuity at the origin $\tau = |z| =0$ , this time of the coefficient $B'(\tau)/B(\tau)$. 
As a consequence we did not obtain a stable numerical solution for the eigensystem of eq.~(\ref{eqxin}).

\section{Conclusions and Outlook}\label{sec:conclusions}

In this paper, we studied the vector and scalar mesons associated with the vector and scalar modes of the 
type IIB supergravity solution, presented in \cite{Dymarsky:2009cm}, corresponding to a stable non-supersymmetric $D7/\overline{D7}$-brane embedding geometrically realizing flavor chiral 
symmetry breaking in the Klebanov-Strassler background. In \cite{Sakai:2003wu}, it proved difficult to find a solution to the equations governing the
 background gauge field. Therefore the authors set $A_{\alpha}=0, \, \alpha =5,6,7$, even though this is not allowed since the background gauge field
 couples to a non-trivial current. A resolution of this problem was presented in \cite{Dymarsky:2009cm} (see discussion above), enabling us to 
study all relevant vector and scalar fluctuation modes (at least for the antipodal embedding $\tau_0=0$), including the pseudo scalar mode associated
 with the non-trivial background gauge field $A_7(\tau)$. 

We found a consistent set of eigenfunctions for the vector mesons, axial vector mesons and the pion. From the corresponding eigenvalues we estimated the 
masses and found a reasonable agreement with previous models and with experiment. For the scalar fluctuations we calculated the effective action from 
 the DBI-WZ expansion but didn't find a singlet combination that leads to a consistent set of eigenfunctions. 

The meson wave functions that we obtain in this model are only approximately parity symmetric in the radial coordinate 
$z= \pm \tau$. The parity symmetry in the $z$-direction is broken, due to the presence of a non-trivial background gauge field that turns out to be asymmetric in the $z$-coordinate. This issue deserves some further elaboration:\\
The charge C and parity P symmetries of the five-dimensional fields are inherited from the 7+1 dimensional DBI action for the $D7$-branes. After Kaluza-Klein reduction on $S^3$, they become the five-dimensional symmetries $P: (x^{\mu},z) \rightarrow (-x^{\mu}, -z)$ and $C: A_{\mu} (x^{\mu},z) \rightarrow -A^T_{\mu} (-x^{\mu},-z)$.
The classical configuration that minimizes the 7+1 dimensional action (i.e., the $D7/\overline{D7}$-embedding together with an ASD/SD gauge field on its world volume) breaks the $z\rightarrow -z$ parity symmetry. However, it is important to observe that parity symmetry is restored asymptotically at large $z$. Therefore we can define the parity quantum numbers of the vector and axial-vector mesons according to their coupling to external gauge fields 
$V= A_L + A_R$ (representing a photon) and $A= A_L- A_R$, resp., at the boundary $|z|\rightarrow \infty$.\\
It seems that the asymmetry of the wave functions in the $z$-direction could be related to a violation of charge conjugation in 3+1 dimensions, but we leave the precise physical interpretation to future research.

It would be very interesting to study the meson spectra for general $\tau_0 > 0$ in order to be able to study more realistic particle phenomenology. However it seems hard to solve the field equations for the background gauge fields even numerically in that case.  We leave this as an open problem for future study. Some of the present authors are currently investigating baryons in this setup, which is also relevant phenomenologically, e.g., for studying nuclear physics in the context of this holographic model. The baryons in this context are usually modelled as instantons, where the instanton number is interpreted as the baryon number. It seems promising that, unlike in the Sakai-Sugimoto model,  the baryon size does not go to zero as $\lambda^{-1/2}$ in the large t'Hooft coupling limit $\lambda \rightarrow \infty$, which is related to the fact that the five-dimensional effective $\kappa$ does 
not depend on $\lambda$. It would also be of considerable interest to study in detail other phenomenological questions such as pion and vector meson form factors and deep inelastic scattering of pion and mesons in the context of this model.\\
Another important aspect in the context of holographic QCD is the introduction of holographic baryons to study, e.g., nucleons and nuclear forces in holographic models. For a more recent discussion of holographic baryons in the DKS model, cf. \cite{Dymarsky:2010ci}. This line of work is based on original ideas by E. Witten \cite{Witten:1998xy} and has been pursued by many authors in the Sakai-Sugimoto $D4/D8$ model \cite{Hata:2007mb,Hashimoto:2008zw,Seki:2008mu, Hashimoto:2009ys}. The conclusion of a recent article \cite{Kaplunovsky:2010eh} was that there cannot be a `realistic' combination of attractive and repulsive nuclear forces in the Sakai-Sugimoto model. As explained in \cite{Kaplunovsky:2010eh}, in order to get an attractive nuclear force slightly greater than the repulsive nuclear force, 
\begin{equation}
V^{\text{rep.}}(r) \propto + r^{-1} e^{-r M_{\text{gs}, \text{vec.}}}, \quad V^{\text{attr.}}(r) \propto - r^{-1} e^{-r M_{\text{gs}, \text{sc.}}}.
\end{equation}
one needs the lightest isoscalar scalar meson to be lighter than the lightest vector meson. Unfortunately, we have not managed to find a stable scalar meson spectrum in our setup for $\tau_0=0$. The situation may be different at $\tau_0 \neq 0$ and especially in the limit $\tau_0 \rightarrow \infty$ where we expect the relevant equations to simplify significantly.
Therefore, we can not yet directly address the question of the mass of the $\sigma$-meson compared to the $\omega$-meson.

In conclusion, it would be worthwhile to study in detail the non-antipodal embedding in the DKS model (for $\tau_0>0$), to see if it can provide more realistic physics. While some phenomenological properties of the DKS model look very promising, more work must be done to see if the model can address phenomenological questions in QCD successfully.

\section*{Acknowledgements}
The authors are partially supported by CAPES and CNPq (Brazilian funding agencies). C.A.B.B. would like to acknowledge the support of the STFC Rolling Grant ST/G000433/1. We would like to thank Jacob Sonnenschein and Anatoly Dymarsky for many helpful comments and discussions. 

\appendix
\section{The deformed conifold}\label{ap:A}
The deformed conifold is a non-singular deformation of the singular conifold:
\begin{equation}
 \sum_{i=1}^4 z_i^2 = -2 \mathrm{det}_{i,j} W = \eps^2,
\end{equation}
where 
\begin{equation}
W= \left( \begin{array}{cc} w_{11} & w_{12} \\ w_{21} & w_{22} \end{array}\right) = \frac{1}{\sqrt{2}} \left( \begin{array}{cc} z_3 + i z_4 & z_1 - i z_2 \\ z_1 +i z_2 & -z_3 + i z_4 \end{array}\right).
\end{equation}
In terms of the radius $\rho$ of the $S^3$, $\mathrm{Tr} W^{\dag} W =\rho^2$, one can define a new coordinate,
\begin{equation}
\rho^2 = \eps^2 \cosh \tau. 
\end{equation}
This is useful, because $\tau =0$ at the tip of the deformed conifold, where the $S^2$ shrinks to zero.
There are many different coordinate bases used to parametrize the deformed conifold in the literature. The reader is referred to the appendices of the review article \cite{Gwyn:2007qf} for a detailled discussion of the different coordinate systems, including some of their shortcomings.
Klebanov and Strassler \cite{Klebanov:2000hb} use the diagonal basis originally introduced by \cite{Candelas:1989js,Minasian:1999tt}. 
The metric reads
\begin{eqnarray}
&&\eps^{-4/3} \d s_{(6)}^ 2 = \\\nonumber
&& \frac{K(\tau)}{2}  \left(\frac{1}{3 K(\tau)^3 } (\d \tau^2 + (g^5)^2) + \cosh^2 \left(\frac{\tau}{2}\right) ((g^3)^2+ (g^4)^2) + \sinh^2 \left(\frac{\tau}{2}\right) ((g^1)^2 + (g^2)^2)\right),
\end{eqnarray}
where 
\begin{equation}
 K (\tau) = \frac{(\sinh (2\tau) - 2 \tau)^{1/3}}{2^{1/3} \sinh \tau}, 
\end{equation}
and the various R-R and NS-NS forms can be found, e.g., in \cite{Klebanov:2000hb}.
For large $\tau$, in terms of the radial coordinate $r \sim \rho^{2/3}$, the deformed conifold metric approaches the metric of the singular conifold
\begin{equation}
\d s_{(6)}^2 \rightarrow \d r^2 + r^2 \d s_{T^{1,1}}^2. 
\end{equation}
It was argued in \cite{Gwyn:2007qf}, however, that this choice of vielbeins does not allow for the construction of a closed holomorphic three-form in the deformed conifold case, implying problems with the Calabi-Yau condition.  
Here, we will make use of the one-forms $e_i$ on $S^2$ and $\e_j$ on $S^3$ that were introduced by \cite{Papadopoulos:2000gj, Butti:2004pk} and used by Dymarsky, Kuperstein and Sonnenschein \cite{Dymarsky:2009cm} to write the metric (\ref{eq:6dmetric}). We utilize this basis in the main part of the paper. This basis is non-diagonal and one can observe that the $S^3$ is fibered over the $S^2$ explicitly:
\begin{eqnarray}
 e_1 &=& \d \theta_1,\nonumber\\
 e_2 &=&\sin \theta_1 \d \phi_1,\nonumber\\
 \e_1 &=& \cos \psi \d \theta_2 + \sin \psi \sin \theta_2 \d \phi_2 = \e_1,\nonumber\\ 
 \e_2 &=& - \sin \psi \d \theta_2 + \cos \psi \sin \theta_2 \d \phi_2,\nonumber\\ 
 \widetilde{\e}_3 &=& \d \psi + \cos \theta_1 \d \phi_1 + \cos \theta_2 \d \phi_2 = \e_3 + \cos \theta_1 \d \phi_1.\nonumber
\end{eqnarray}
Since the non-BPS $D7$-brane embedding described in \cite{Dymarsky:2009cm} covers the $S^3$ completely, 
i.e., preserves the $SU(2)_R$ invariance of the deformed conifold, it is not necessary for our purposes to use a basis in which the coordinates on $S^3$ are independent of the coordinates on $S^2$. \\
Such a trivialization basis was found in \cite{Evslin:2007ux} and written down explicitly in \cite{Krishnan:2008gx} (see also \cite{Gimon:2002nr}). It is stated here for completeness and because it may be useful in other contexts. In this $S^3 \times S^2$ basis, the metric becomes
\begin{eqnarray}
 \eps^{-4/3} \d s_{(6)}^ 2&=&\frac{1}{6K^2(\tau)} (\d \tau^2 + h_3^2) + \frac{K(\tau)}{4} \cosh^2 \left(\frac{\tau}{2}\right) \times \nonumber\\  
 &\times& \left[ h_1^2 + h_2^2 + 4 \tanh^2 \left(\frac{\tau}{2}\right) \left(\left(\d \theta -\frac{1}{2} h_2\right)^2 +\left(\sin \theta \d \phi - \frac{1}{2} h_1\right)^2\right)\right].
\end{eqnarray}
The one-forms $h_i$, $i=1,2,3$ manifest the fibration of the $S^3$ over the $S^2$:
\begin{equation}
 \left( \begin{array}{c} h_1 \\ h_2 \\ h_3 \end{array} \right)= \left(\begin{array}{ccc} 0 & \cos \theta & - \sin \theta \\ 1 & 0 & 0 \\
0 & \sin \theta & \cos \theta \end{array}\right) \left(\begin{array}{ccc} \sin \phi & \cos \phi & 0 \\ \cos \phi & -\sin \phi & 0 \\
0 & 0 & 1 \end{array}\right)  \left( \begin{array}{c} w_1 \\ w_2 \\ w_3 \end{array} \right).
\end{equation}
The function $K(\tau)$ is identical to the one defined above and the $SU(2)$ left-invariant Maurer-Cartan forms $w_i$, $i=1,2,3$ read
\begin{eqnarray}
-\frac{w_1}{2} &=& -\cos \beta \mathrm{d} \alpha + \sin \alpha \cos \alpha \sin \beta \mathrm{d} \beta -\sin^2 \alpha \sin^2 \beta \mathrm{d} \gamma,\\ 
-\frac{w_2}{2} &=& -\sin \beta \cos \gamma \mathrm{d} \alpha - (\sin \alpha \cos \alpha \cos \beta \cos \gamma - \sin^2 \alpha \sin \gamma) \mathrm{d} \beta \nonumber \\ &&+ (\sin^2 \alpha \sin \beta \cos \beta \cos \gamma + \sin \alpha \cos \alpha \sin \beta \sin \gamma) \mathrm{d} \gamma,\\
\frac{w_3}{2} &=& \sin \beta \sin \gamma \mathrm{d} \alpha + (\sin \alpha \cos \alpha \cos \beta \sin \gamma + \sin^2 \alpha \cos \gamma)\mathrm{d}\beta \nonumber \\ && + (\sin \alpha \cos \alpha \sin \beta \cos \gamma - \sin^2 \alpha \sin \beta \cos \beta \sin \gamma) \mathrm{d} \gamma,
\end{eqnarray}
where $\alpha,\beta,\gamma$ are coordinates on $S^3$.

\section{Expansion of the DBI action}\label{ap:B}
Consider the pullback of the 10-d coordinates $X^M=(x^\mu, \tau, \theta_1, \phi_1, \Omega_i)$
 on the D7 brane worldvolume with coordinates $X^\alpha = (x^\mu, \tau, \Omega_i)$ where $\theta_1 = \theta_1(\tau, x)$, $\phi_1=\phi_1(\tau,x)$ and $d \Omega_i = \epsilon_i$ ($i=1,2,3$) 
 
 The pullback is defined by 
 \beq
(g_{(8)})_{M N} = G_{P Q} \partial_M X^P \partial_N X^Q
 \eeq
 where $G_{PQ}$ is the 10-d metric defined by 
 \beqa
ds^2 &=& h^{-1/2} \eta_{\mu \nu} dx^\mu dx^\nu  + h^{1/2} \frac{\epsilon^{4/3}}{4} K \cosh \tau \Big \{ d \theta_1^2 + \sin^2 \theta_1 d\phi_1^2  + d\Omega_1^2 + d\Omega_2^2     \cr 
&+& \frac{2}{\cosh \tau} ( d \theta_1 d\Omega_1 + \sin \theta_1 d \phi_1 d \Omega_2 ) 
+ \frac{2}{3 K^3 \cosh\tau}[ d\tau^2 + (d\Omega_3 + \cos \theta_1 d \phi_1)^2 ] \Big \} 
 \eeqa
For the antipodal case the pullback takes the form 
\beqa
(g_{(8)})_{MN} = ( h^{-1/2} \eta_{\mu \nu} , h^{1/2} g_{\alpha \beta} )  \, ,
\eeqa
where $\alpha =(\tau,1,2,3)$ and
\beqa
g_{\alpha \beta} = \frac{\epsilon^{4/3}}{4} K \cosh \tau \, \, {\rm diag} \left(\frac{2}{3 K^3 \cosh \tau}, 1, 1,  \frac{2}{3 K^3 \cosh \tau}\right) \,. 
\eeqa

On the other hand, the gauge-invariant field strength is
\beqa
{\cal F} &=& P [ B_{(2)} ] + F \cr
&=& \left ( -  \frac{b \partial_\tau \phi_1 }{\cosh \tau}  + \partial_\tau A_5 \right ) d \tau \wedge \epsilon_1 + \partial_\tau A_6 \,d \tau \wedge \epsilon_2  + \partial_\tau A_7 \,d \tau \wedge \epsilon_3 \cr 
&-& A_5 \,\epsilon_2 \wedge \epsilon_3 + A_6 \,\epsilon_1 \wedge \epsilon_3 + (b - A_7) \epsilon_1 \wedge \epsilon_2  \,+\, \cr 
&+& \partial_\mu A_5 \,d x^\mu \wedge \epsilon_1 + \partial_\mu A_6 \,dx^\mu \wedge \epsilon_2  + \partial_\mu A_7 \,d x^\mu \wedge \epsilon_3 \,+\, \frac12 F_{\mu \nu} dx^\mu \wedge dx^\nu \cr 
&=: & \frac12 {\cal F}_{MN} \, d X^{M}\wedge d X^{N} 
\eeqa
where $2\pi \alpha'$ has been absorbed in the gauge fields. In the antipodal case it takes the form 
\beqa
{\cal F}_{M N} = ( 0 , {\cal F}_{\alpha \beta} ) \, ,
\eeqa
with
\beq
 {\cal F}_{12} = b - A_7\quad , \quad 
{\cal F}_{\tau 1} = {\cal F}_{\tau 2} = {\cal F}_{1 3} = {\cal F}_{1 3} = 0 \quad , \quad 
 {\cal F}_{\tau 3} = -\frac{g_{\tau \tau}}{g_{11}} {\cal F}_{1 2} \, , 
\eeq

The DBI action can be written as 
\beq
S_{DBI} = - \mu_7 \int d^4 x d \tau d^3 \Omega e^{- \Phi} \sqrt{| \det ( E_{MN} + \delta E_{MN} )|} \, .
\eeq
where 
\beqa
E_{MN}= (g_{(8)})_{MN} + {\cal F}_{MN} \, , 
\eeqa
is the background matrix and 
\beqa
\delta E_{MN} = \delta(g_{(8)})_{MN} + \delta{\cal F}_{MN}
\eeqa
the fluctuation matrix. The background matrix has an inverse given by 
\beqa
E^{MN} = \left ( h^{1/2} \eta^{\mu \nu} ,  E^{\alpha \beta} \right )
\eeqa
where
\beqa
E^{\alpha \beta} = \frac{h^{-1/2} \sqrt{ |g|}}{\sqrt{|E|}}(g^{-1} - h^{-1/2} g^{-1} {\cal F} g^{-1} )^{\alpha \beta} 
\eeqa
and the square root of its determinant can be written as
\beq
\sqrt{|E|} = \sqrt{|g|} +h^{-1} \sqrt{|{\cal F}|} = \sqrt{\frac{g_{\tau \tau} g_{33}}{g_{11}g_{22}}} h^{-1} \left ( {\cal F}_{1 2}^2  + g_{11}g_{22} h \right) 
\eeq

We omit for a moment the tensor notation. We want to expand the determinant 
\beq
\det (E + \delta E) = \det (E) \det ( 1 + E^{-1} \delta E ) \, ,
\eeq
Using the identity 
\beq
\det (1 + X) =   \exp \{ \tr [ \ln (1 + X ) ] \} 
\eeq
and the logarithm expansion 
\beq
\ln (1 + X ) = X - \frac12 X^2 + \dots , 
\eeq
we obtain 
\beqa
\det (E + \delta E) &=&  \det (E) \Big \{ 1 + \tr [ E^{-1} \delta E] + \frac12 ( \tr [ E^{-1} \delta E])^2  \cr 
&-& \frac12 \tr  [(E^{-1} \delta E)^2] + {\cal O} ((\delta E)^3 ) \Big \} \, ,
\eeqa
so that 
\beqa
\sqrt{-\det (E + \delta E)} &=& \sqrt{-\det(E)} \Big[ 1 + \frac12 \tr [ E^{-1} \delta E ] 
+ \frac18 (\tr [ E^{-1} \delta E ])^2 \cr 
&-& \frac14 \tr [ (E^{-1} \delta E)^2 ] + {\cal O} ((\delta E)^3 ) \Big ]\, . 
\eeqa

\end{document}